\documentclass[print]{aastex61}

\usepackage{hyperref}
\usepackage{comment}

\received{28 February 2018}
\revised{22 April 2018}
\accepted{23 April 2018}

\submitjournal{ApJ}

\begin{document}

\title{Measuring the Local ISM Along the Sight Lines of the Two \textit{Voyager} Spacecraft with \textit{HST}/STIS}

\author[0000-0002-3693-6770]{Julia Zachary}
\affiliation{Wesleyan University Astronomy Department, Van Vleck Observatory, Middletown, CT 06459-0123, USA; \href{mailto:jzachary@wesleyan.edu}{jzachary@wesleyan.edu}}

\author[0000-0003-3786-3486]{Seth Redfield}
\affiliation{Wesleyan University Astronomy Department, Van Vleck Observatory, Middletown, CT 06459-0123, USA; \href{mailto:jzachary@wesleyan.edu}{jzachary@wesleyan.edu}}

\author[0000-0003-4446-3181]{Jeffrey L. Linsky}
\affiliation{JILA, University of Colorado and NIST, Boulder, CO 80309-0440, USA; }

\author[0000-0002-4998-0893]{Brian E. Wood}
\affiliation{Naval Research Laboratory, Space Science Division, Washington, DC 20375, USA; }

\begin{abstract}

In 2012, \textit{Voyager 1} crossed the heliopause, becoming the first human-made object to exit the solar system. This milestone signifies the beginning of an important new era for local interstellar medium (LISM) exploration. We present measurements of the structure and composition of the LISM in the immediate path of the \textit{Voyager} spacecraft by using high-resolution \textit{Hubble Space Telescope} (\textit{HST}) spectra of nearby stars that lie along the same lines of sight. We provide a comprehensive inventory of LISM absorption in the near-ultraviolet (2600--2800 $\textrm{\AA}$) and far-ultraviolet (1200--1500 $\textrm{\AA}$). The LISM absorption profiles are used to make comparisons between each pair of closely spaced ($<$$15^\circ$) sight lines. With fits to several absorption lines, we make measurements of the physical properties of the LISM. We estimate electron density along the \textit{Voyager 2} sight line, and our values are consistent with recent measurements by \textit{Voyager 1}. Excess absorption in the \ion{H}{1} Ly$\alpha$ line displays the presence of both the heliosphere and an astrosphere around GJ 780. This is only the 14th detection of an astrosphere, and the large mass-loss rate ($\dot{M} = 10\dot{M}_\odot$) is consistent with other subgiant stars. The heliospheric absorption matches the predicted strength for a sight line $58^\circ$ from the upwind direction. As both \textit{HST} and \textit{Voyager} reach the end of their lifetimes, we have the opportunity to synthesize their respective observations, combining in situ measurements with the shortest possible line-of-sight measurements to study the Galactic ISM surrounding the Sun.
    
\end{abstract}


\section{Introduction} \label{sec:intro}

The local interstellar medium (LISM) forms the outer boundary for the heliosphere, and its properties dictate the behavior of the Sun and heliosphere \citep{Redfield2004}. The Sun resides within the Local Cavity, a region of low-density material extending to $\sim$100 pc within the local galactic neighborhood \citep{Welsh2010}. Much of the interstellar medium (ISM) consists of low-density, partially ionized gas coupled with dust. The ionization state of the gas depends on the local intensity of ultraviolet radiation needed to photodissociate or photoionize molecules \citep{Draine2009}. Extreme-ultraviolet light from hot stars or white dwarfs provides sources of photoionization, causing an anisotropic gradient within the Local Cavity \citep{Frisch2011}. Structured LISM clouds were only identified 50 yr ago, and the first spectrum of local interstellar gas outside the heliosphere was taken in 1977 with the \textit{Copernicus} satellite \citep{Adams1977, Frisch2009}. Understanding the physical properties of the LISM is critical to understanding the current structure and evolution of the heliosphere.

The kinematic structure of the LISM is moderately complicated. With an average of 1.5 absorbers per LISM sight line \citep{RedfieldLinsky2004a}, the richness of the kinematic structure can be evaluated by resolving the LISM absorption profiles. More distant sight lines are difficult to decipher given the extreme blending of ISM components. Early analysis of titanium absorption-line spectra for stars within 100 pc by \citet{Crutcher1982} found that warm gas within the LISM moves with a coherent heliospheric velocity. The kinematic model of the LISM by \citet{RedfieldLinsky2008} includes 15 clouds moving with distinct but similar bulk velocity vectors. These component clouds are individual comoving structures of partially ionized gas identified by common velocities across large patches of the sky \citep{RedfieldLinsky2015}. Most LISM clouds consist of warm to hot neutral or ionized gas maintained by photoionization. Analysis of \textit{Ulysses} mission data by \cite{Wood2015} found helium temperatures of 7260$\pm$270 K for gas penetrating the heliosphere, and line-of-sight observations to nearby stars yielded measurements of LISM cloud temperatures between 5000 and 8000 K \citep{RedfieldLinsky2004b}. 

\textit{ROSAT} soft x-ray observations show that the partially ionized LISM clouds are immersed in a hot ($\lesssim$10$^6$ K) medium \citep{Galeazzi2014}. \cite{Slavin2014} noted that thermal conduction between the surrounding hot gas and the warm local clouds leads to evaporation. In order for the cloudlets to survive, the thermal and magnetic pressure of the clouds must equal the thermal pressure of the Local Hot Bubble to maintain pressure equilibrium \citep{Slavin2014}. This is supported by \cite{Frisch2011}, who determine that the thermal pressure of the hot gas in the Local Cavity is in the range of $\rho/k \sim 3000-7000$ cm$^{-3}$ K to be in pressure equilibrium. Spectroscopic data that sample the wavelength range 1540$-$1555 $\textrm{\AA}$ contain the stellar and interstellar \ion{C}{4} absorption lines, which can be used as an indication of thermal conduction. Observations of \ion{C}{4} by \textit{Hubble Space Telescope} (\textit{HST}) and \textit{FUSE} give much smaller column densities than predicted by cloud evaporation models \citep{Slavin2004}. Therefore, it is uncertain that LISM clouds are evaporating.

The properties of LISM dust have been inferred from extinction and polarization of starlight, scattering, and thermal emission \citep{Draine2009}. Interstellar clouds in the LISM are expected to be diffuse, consisting largely of neutral hydrogen with low extinction. Spectroscopic features observed in the infrared show silicate absorption, consistent with the assumption that most interstellar dust is composed of silicates or carbonates \citep{Draine2009}. The number of atoms that must be depleted onto dust grains can also be calculated based on pickup ion isotope ratios \citep{Frisch2011}. Elements with the highest condensation temperatures are the most depleted in the ISM, and the gas phase abundances of volatile elements are higher in warm clouds \citep{Savage1996, Frisch2011}. Small depletions are well correlated with high turbulent velocities, suggesting that the destruction of dust grains may return specific ions to their gas phase. Dust destruction may also be caused by shocks produced by supernovae, by turbulent motions driven by interactions between clouds, or by direct collisions \citep{RedfieldLinsky2008}. 

One common way of studying the LISM is to observe absorption features against nearby stellar background sources. The most important resonance lines of ISM ions are found at UV wavelengths. The \textit{HST} Space Telescope Imaging Spectrograph (STIS) is a general-purpose spectrograph capable of observing across a wide range of wavelengths and a large number of diverse astrophysical targets. The high spectral resolution capabilities of STIS are ideal for observing LISM absorption-line profiles and resolving the velocity components of LISM clouds. LISM absorption-line profiles are intrinsically narrow, and the velocity components of LISM clouds may overlap. With sufficient spectral resolution to resolve individual velocity components, we can make accurate physical measurements of the temperature, turbulent velocity, and electron density. Short sight lines, like those utilized in this work, permit detailed study of warm LISM material since absorption profiles in spectral lines are less likely to be as blended or saturated compared to long ($\sim$100$-$1000 pc) sight lines \citep{Malamut2014}. Observing heavy ions (e.g., \ion{Mg}{2} and \ion{Fe}{2}) with STIS provides information about the kinematic structure of the LISM because these ions are less impacted by thermal broadening and blending. We can thereby easily identify multiple ISM components along a particular line of sight.

On 1977 September 5, \textit{Voyager 1} was launched from the NASA Kennedy Space Center at Cape Canaveral, Florida.\textit{Voyager 1} and \textit{Voyager 2}, which was launched in 1977 August, studied the interplanetary space between Earth and Saturn and explored the Saturnian and Jovian planetary systems. \textit{Voyager 2} successfully extended its mission to Uranus and Neptune. By the end of 1989, both spacecraft headed on separate paths out of the ecliptic plane. At that point NASA initiated the \textit{Voyager} Interstellar Mission. Its goals are not only to continue investigating the interplanetary medium but also to characterize the structure of the heliosphere and ultimately to study the ISM \citep{Rudd1997}. 

The spacecraft have now endured over 40 yr in space, and their radioisotope thermoelectric generators will continue to power onboard instruments until 2025--2030. Out of the 11 scientific instruments originally operational on the \textit{Voyagers}, only five remain in use on board \textit{Voyager 2} and four on board \textit{Voyager 1}. These instruments are crucial to providing insight into the composition, structure, and presence of the various components of the heliosphere and interstellar space. As of 2018 April 22, \textit{Voyager 1} and \textit{Voyager 2} are 141.3 and 117.2 au from Earth, respectively.\footnote{https://voyager.jpl.nasa.gov/mission/status/} \textit{Voyager 1} is already in the LISM, having crossed the termination shock in 2004 at 94 au \citep{Decker2005} and the heliopause in 2012 \citep{Gurnett2013}. \textit{Voyager 2} remains in the heliosheath, having crossed the termination shock in 2007 at 84 au \citep{Richardson2008}. As the \textit{Voyager} spacecraft move toward pristine LISM environments, it would be interesting to compare the complimentary complimentary in situ observations with line-of-sight measurements.

We present work that is able to take advantage of the advanced capabilities of STIS by using high resolution spectra of four nearby stars along the sight lines of the \textit{Voyager} spacecraft to compare with {\em in situ} data from the spacecraft themselves. The spectra provide a host of quantitative measurements of nearby ISM gas. By observing stars within 15$^\circ$ of the lines of sight of the \textit{Voyagers}, we create a direct connection to the in situ measurements taken by the spacecraft. Analysis of the \textit{HST} spectra in the context of past LISM studies allows us to provide a far larger overview, and we can predict what kind of ISM environment the \textit{Voyagers} may one day travel through.

\section{Observations} \label{sec:obs}

The spectra for this work were taken with STIS in either the near-UV (NUV) or far-UV (FUV) between 1150 and 3100 $\textrm{\AA}$ using the NUV and FUV MAMA detectors. The observations utilized three of the four higher-order echelle gratings: E230H, E140M, and E140H with resolving power $R = {\lambda\over\Delta\lambda}\sim$ 114,000, 45,800, and 114,000, respectively. The E230H grating was used with the NUV MAMA detector to provide spectra over the $2576-2812 \textrm{\AA}$ wavelength range. We selected this grating to observe the \ion{Mg}{2} ($\lambda\lambda$2796, 2803) and \ion{Fe}{2} ($\lambda\lambda$2586, 2600) ions, which are not significantly thermally broadened, but provide sharp line profiles to resolve line of sight velocity structure. We used the E140M grating to observe the Ly$\alpha$ line of \ion{H}{1} ($\lambda$1215), which has a broad interstellar absorption profile in its line core. This setting was used with the FUV MAMA detector, providing a wavelength coverage of $1144-1710 \textrm{\AA}$. The E140H grating has several different settings, each extending over small, separate wavelength ranges. For our brightest target (GJ 780), we used an E140H setting that included wavelengths between 1176 and 1372 $\textrm{\AA}$ to obtain the Ly$\alpha$ line of \ion{H}{1}. The E140M grating utilized the $0.2\times0.2$ aperture, while the E230H and E140H gratings used the $0.2\times0.09$ aperture \citep{Woodgate}.

We obtained data during four nonconsecutive visits between 2015 August and October. Each individual observation was devoted to one target star with observations taken over the course of 4--8 hr. A full table of observation parameters is shown in Table 1.

\begin{deluxetable*}{clccccccc}[]
\tablecolumns{9}
\tablewidth{0pc}
\tablecaption{Observational Parameters for the \textit{HST}/STIS Data Used in This Work.}
    \tablehead{
    \colhead{Target} & \colhead{Data Set} & \colhead{Observation} & \colhead{Exposure} & \colhead{Aperture} & \colhead{Filter/} & \colhead{Wavelength} & \colhead{S/N} & \colhead{S/N }\\
     \colhead{Name} & \colhead{} & \colhead{Date} & \colhead{Time} & \colhead{} & \colhead{Grating} & \colhead{Range} & \colhead{\ion{Mg}{2}} &  \colhead{Ly$\alpha$}\\
      &  &  &  (s) & (arcsec $\times$ arcsec) & &  ($\textrm{\AA}$) &  (2796 $\textrm{\AA}$) &  (1215 $\textrm{\AA}$)}
    \startdata
    GJ 754 & OCMN04010 & 2015 Oct 1 & 1751 & $0.2\times0.09$ & E230H & $2576-2812$ & 4.5 & \nodata\\
    & OCMN04020--50 & & 3106 & $0.2\times0.2$ & E140M & $1144-1710$ &\nodata & 12.5\\
    GJ 780 & OCMN03010 & 2015 Sep 19 & 2185 & $0.2\times0.09$ & E230H & $2576-2812$ & 81.0& \nodata\\
    & OCMN03020--30 & & 3320 & $0.2\times0.09$ & E140H & $1176-1372$ &\nodata & 19.5\\
    GJ 686 & OCMN02010 & 2015 Sep 24 & 1871 & $0.2\times0.09$ & E230H & $2576-2812$ & 7.0& \nodata\\
    & OCMN02020--30 & & 2979 & $0.2\times0.2$ & E140M & $1144-1710$ &\nodata & 12.0\\
    GJ 678.1A & OCMN01010 & 2015 Aug 15 & 1868 & $0.2\times0.09$ & E230H & $2576-2812$ & 13.6& \nodata\\
    & OCMN01020--30 & & 2970 & $0.2\times0.2$ & E140M & $1144-1710$ &\nodata & 9.0\\
    \enddata
\end{deluxetable*}

\subsection{Target Stars}
We selected target stars for observations with \textit{HST}/STIS on the basis of their close angular separations from the \textit{Voyager} lines of sight. Prior to these observations, there were no spectra of any nearby ($<$10 pc) stars within $\Delta\theta = 15^\circ$ of the \textit{Voyager} spacecraft trajectories. At the time when the observations with STIS were taken, \textit{Voyager 1} was traveling out of the ecliptic plane with $\lambda = 255^\circ.40$, $\beta = 35^\circ.01$, $\ell = 32^\circ.75$, and $b = 28^\circ.08$ toward the constellation Ophiuchus. \textit{Voyager 2} was traveling out of the ecliptic plane with $\lambda=289^\circ.76$, $\beta = -35^\circ.46$, $\ell=340^\circ.55$, and $b = -31^\circ.35$ toward the constellations of Sagittarius and Pavo.\footnote{https://ssd.jpl.nasa.gov/horizons.cgi} The use of two targets per sight line permits us to identify small-scale structure because short sight lines yield simple absorption profiles \citep{Malamut2014}. Additionally, two targets per sight line provided some redundancy in case we did not detect any LISM absorption toward one star. Our targets consist of three M dwarfs (GJ 678.1A, GJ 686, and GJ 754) and one late G-type star (GJ 780). GJ 780 has $V = 4.62$ mag and yields the highest signal-to-noise ratio (S/N) out of all the spectra we obtain. All four targets are located within 10 pc of the Sun, thereby allowing for close study of the LISM. The stellar parameters and angles from the \textit{Voyager} sight lines are presented in Table 2.

\begin{deluxetable*}{cccccccccccc}[]
\tablecolumns{9}
\tablewidth{0pc} 
\tablecaption{Stellar Parameters for the Target Stars along the Lines of Sight toward the \textit{Voyager} Spacecraft.}
\tablehead{
\colhead{Gliese} & \colhead{Other} & \colhead{Spectral} & \colhead{$l$} & \colhead{$b$} & \colhead{$v_{\mathrm{radial}}$} & \colhead{$V$} & \colhead{$\Delta\theta$} & \colhead{$d$}\\
\colhead{No.} & \colhead{Name} & \colhead{Type} & (deg) & (deg) & (km s$^{-1}$) & (mag) & (deg) & (pc)
}
\startdata
\multicolumn{9}{c}{\textit{Voyager 1}}\\
\cline{1-9}
678.1A & HIP 85665 & M1.0Ve\tablenotemark{a} & 028.57 & +20.54 & $-12.51$\tablenotemark{b} & 9.43\tablenotemark{c} & 8.1 & $9.98\pm0.11$\tablenotemark{d}\\
686 & HIP 86287 & M1.5Ve\tablenotemark{a} & 042.24 & +24.30 & $-9.55$\tablenotemark{b} & 9.62\tablenotemark{e} & 9.0 & $8.09\pm0.11$\tablenotemark{d}\\
\cline{1-9}
\multicolumn{9}{c}{\textit{Voyager 2}}\\
\cline{1-9}
780 & $\delta$ Pav & G8IV\tablenotemark{f} & 329.77 & $-32.42$ & $-21.7$\tablenotemark{g} & 4.62\tablenotemark{h} & 9.2 & $6.11\pm0.03$\tablenotemark{i}\\
754 & LHS 60 & M4V\tablenotemark{j} & 352.36 & $-23.90$ & 16.0\tablenotemark{k} & 12.25\tablenotemark{l} & 13.1 & $5.92\pm0.05$\tablenotemark{m}\\
\enddata
\tablecomments{Galactic coordinates from the SIMBAD Astronomical Database.
\\$^\mathrm{a}$\cite{Lepine2013}, $^\mathrm{b}$\cite{Nidever2002}, $^\mathrm{c}$\cite{Zacharias2012}, $^\mathrm{d}$\cite{Koen2010}, $^\mathrm{e}$\cite{Astudillo-Defru2017}, $^\mathrm{f}$\cite{Gray2006}, $^\mathrm{g}$\cite{Evans1979}, $^\mathrm{h}$\cite{Sousa2008}, $^\mathrm{i}$\cite{vanLeeuwen2007}, $^\mathrm{j}$\cite{Bonfils2013}, $^\mathrm{k}$\cite{Rodgers1974}, $^\mathrm{l}$\cite{Winters2015}, $^\mathrm{m}$\cite{Jao2005}.}
\end{deluxetable*}

\subsection{Data Reduction}

We use the Space Telescope Science Institute data reduction pipeline \texttt{calstis},\footnote{http://stsdas.stsci.edu/cgi-bin/gethelp.cgi?calstis} which performs basic 2D image reduction to produce a flat-fielded output image and performs 2D and 1D spectral extraction to produce either a flux-calibrated spectroscopic image or a 1D spectrum of flux versus wavelength \citep{Riley2017}. The \texttt{calstis} pipeline propagates statistical errors and tracks the quality of the data throughout the calibration, flagging data when bad pixels are present. For wavelength calibration, the onboard Pt$-$Cr/Ne calibration lamps were used, followed up by \texttt{calstis} processing the associated wave-calibrated exposure to determine the zero-point offset of the wavelength and spatial scales in the science image \citep{Bristow, Riley2017}.

\section{Fitting} \label{sec:fits}
The first step in the data analysis process is to fit the stellar continuum at the location of the LISM absorption lines. For the cool stars that are used in this work, we observe LISM absorption against stellar line emission \citep{RedfieldLinsky2002}. We fit the intrinsic stellar emission line that serves as a ``continuum'' with a low-order polynomial to the regions both just redward and blueward of the absorption feature. When the interstellar absorption is far from the line center, the unobserved stellar flux
can be estimated by flipping the emission line about the stellar radial velocity \citep{RedfieldLinsky2002}. It is crucial to get the best possible estimate for the intrinsic stellar continuum upon which the interstellar absorption is superimposed, since any systematic uncertainties in this assumed intrinsic line profile propagate to our LISM fit parameters \citep{LinskyWood1996}. 

Each fit begins with one interstellar absorption component unless it is evident from visual inspection of the data that there are multiple absorbing clouds. We analyze each line profile without imposing any constraints on the characterizing parameters. Our fitting routine utilizes a Levenberg$-$Marquardt least-squares algorithm to fit Voigt absorption profiles to the data. The algorithm
requires initial guesses for the Doppler parameter, absorption centroid wavelength, and $\log$ column density and then varies all parameters until achieving a minimum $\chi^2$. We base these initial guesses off both visual inspection of the data and known average values for each ion's Doppler parameter and column density. The instrumental line-spread function for STIS \citep{Riley2017} is also incorporated in our fit. After the initial fits are achieved, we run a Monte Carlo (MC) error analysis to determine the uncertainty on each parameter. Unless noted, the uncertainties listed in Tables 3--4 are those generated by MC error analysis. 

Absorption features often contain more than one cloud component. In order to determine the number of absorbers, we start from one component and increase the number of absorbers as warranted by the data until the quality of the fit improves \citep{RedfieldLinsky2002}. With the addition of each component, the $\chi^2$ value decreases. At a certain point, continuing to increase the number of components no longer significantly improves the quality of the fit as determined by an $F$-test \citep{Bevington2003}. 

Certain ions, such as \ion{Fe}{2} and \ion{Mg}{2}, contain multiple resonance lines in a single wavelength range. Since each
line of a given multiplet must contain the same interstellar absorption components at the same radial velocity with the same line widths and column densities modified by their relative $gf$ values, we can fit all lines in a multiplet simultaneously \citep{RedfieldLinsky2002}. The lines of a doublet provide independent measurements of the same ion, and the difference in oscillator strengths between the components of the doublet provides accurate constraints on ISM absorption parameters, stellar continuum flux, and the total number of absorbers. Simultaneous fits to all lines in a multiplet provide a better determination of interstellar absorption parameters than fits to individual lines, although we perform individual fits for comparison. 

\section{Results} \label{sec:results}
We fit interstellar absorption components for all four sight lines. We first took inventory of the LISM absorption detected in the spectra and fit individual ions, leaving the \ion{H}{1} Ly$\alpha$ analysis until later (Section 4.1). We detect interstellar absorption by \ion{Mg}{2} and \ion{D}{1} in GJ 678.1A (see Figure 1 and Table 3); \ion{D}{1} in GJ 686 (see Figure 1 and Table 3); \ion{Mg}{2}, \ion{O}{1}, \ion{C}{2}, \ion{C}{2}$^*$, and \ion{D}{1} in GJ 780 (see Figure 1 and Table 4); and \ion{D}{1} in GJ 754 (see Figure 1 and Table 4).

The bright star GJ 780 shows evidence of two LISM absorption components in multiple ions and is the most complete set of LISM absorption features. Since both GJ 754 and GJ 686 are faint M stars, the low S/N did not allow us to detect and fit \ion{Mg}{2} ISM absorption along their respective sight lines (see Fig. 2). We note in both stars that there is a large shift in the radial velocity of the star in comparison with the predicted velocities of the ISM clouds, so the ISM absorption is shifted away from the stellar \ion{Mg}{2} lines. Thus, no significant LISM absorption is detected at the predicted velocities, perhaps due to the low observed S/N and large velocity offset that shifted the interstellar absorption away from the bright stellar emission line cores. We detect the Ly$\alpha$ lines of \ion{H}{1} and \ion{D}{1} in all four target stars. The fits to the LISM absorption in the target stars are presented in Figure 1. 

The final parameters for each fit are presented in Tables 3--4. Listed for each component are the heliocentric velocity (\textit{v} [in km s$^{-1}$]), the Doppler parameter (\textit{b} [in km s$^{-1}$]), and the log column density (\textit{$\log_{10} N$} [in cm$^{-2}$]). For \ion{Mg}{2} and \ion{C}{2}, the parameter values are the weighted mean from the individual and simultaneous fits. The associated errors are, in the case of multiple fits, either the standard deviation or the weighted mean errors, or the resulting MC uncertainties if no simultaneous fit was performed. 

\paragraph{GJ 678.1A} In the GJ 678.1A spectra, we determine that there are only two absorption components. We detect both lines of the \ion{Mg}{2} doublet and successfully perform a simultaneous fit. In the \ion{D}{1} line, we froze the Doppler parameters at fixed values in order to construct an accurate fit to the data. When left unfrozen, the $b$ values were too high, and we considered them to be unphysical. 

\paragraph{GJ 686} We find only one absorption component in the \ion{D}{1} Ly$\alpha$ line in GJ 686, shown in Figure 1. We fit the data after the \ion{H}{1} analysis was completed, using the relationship between the \ion{D}{1} and \ion{H}{1} Doppler parameters to constrain the fit.

\paragraph{GJ 780} Figure 1 displays the strong ISM absorption seen in GJ 780. We clearly see two distinct absorption components centered at $-$17 km s$^{-1}$ and $-$9 km s$^{-1}$. We note that both the \ion{C}{2} and \ion{O}{1} absorptions are saturated, although the well-characterized \ion{Mg}{2} LISM component structure allowed us to derive robust measurements of \ion{C}{2} and \ion{O}{1} despite their saturation. We perform a simultaneous fit to the \ion{C}{2} line and its excited state \ion{C}{2}$^*$ as a proxy for measuring the electron density in the ISM (See Section 5.4). Some important LISM transitions (\ion{H}{1} Ly$\alpha$, \ion{O}{1}, \ion{N}{1}) show significant geocoronal contamination from the Earth's exosphere, the outermost and most tenuous layer of Earth's
atmosphere. Beginning at an altitude of 500 km, the exosphere has been detected as far as $\sim$15.5 R$_\odot$ \citep{Schultz2014}. It is composed of mostly neutral hydrogen atoms
and is detected at UV wavelengths as a result of interactions between neutral hydrogen and high-energy solar photons. Geocoronal absorption is readily identified because it is centered at the mean velocity of Earth during the time of observation. The \ion{O}{1} line in GJ 780 (see Fig. 1) shows evidence of geocoronal absorption with velocity $v_r = -19.8$ km s$^{-1}$ at the barycentric velocity of the Earth at the time of our observations. A three-component fit (one geocoronal and two LISM) produces a result consistent with the absorption in other ions. The geocorona is also seen as an emission feature in the hydrogen Ly$\alpha$ line (see Sec. 4.1).

\paragraph{GJ 754} We determine that there are only two absorption components in the Ly$\alpha$ line of \ion{D}{1} in GJ 754, which is shown in Figure 1. \ion{D}{1} absorption features have large Doppler parameters and are indicative of broad absorption features. 

\begin{figure*}
	\centerline{\includegraphics[scale=0.9,trim={0 11cm 0 0},clip]{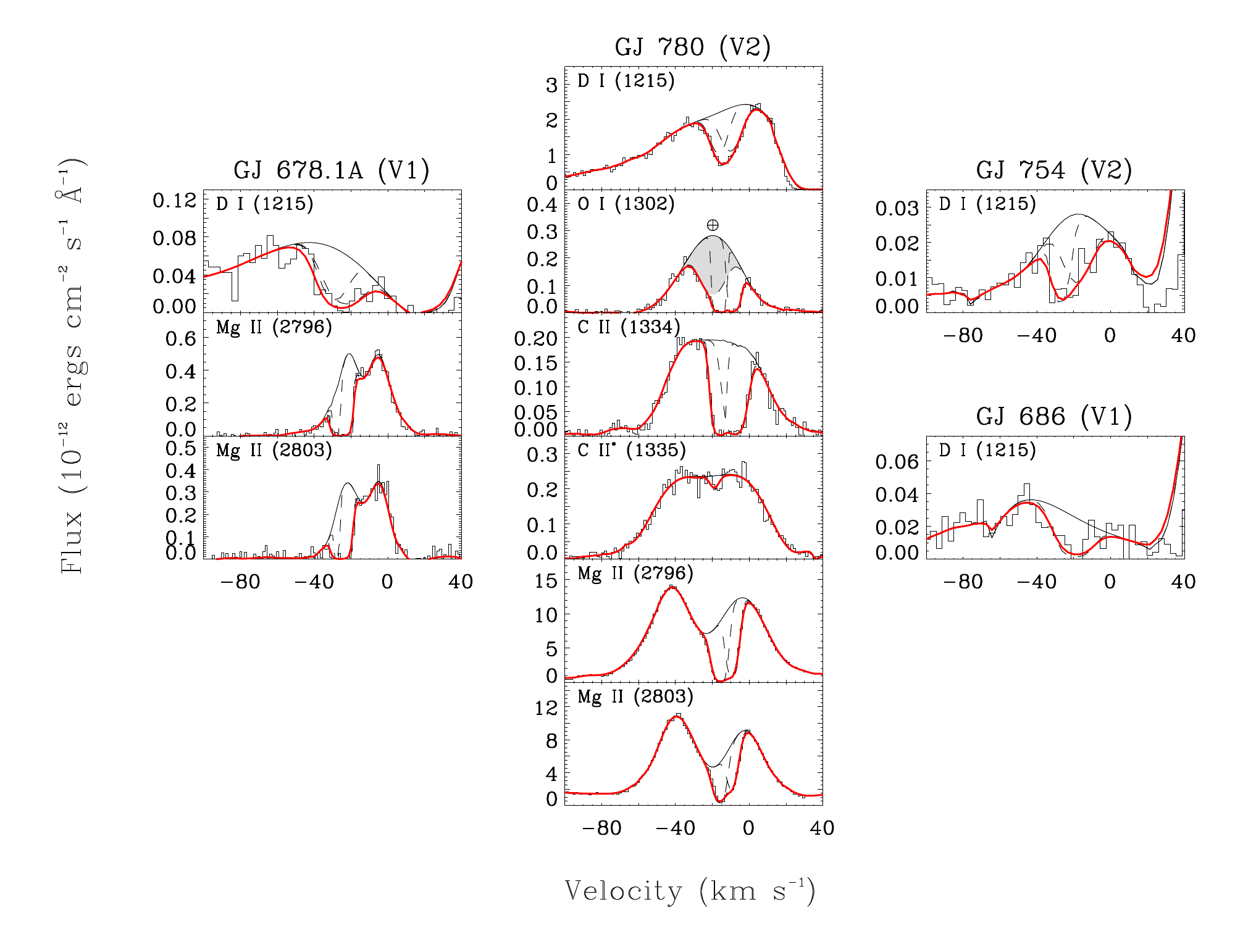}}
	\caption{Best-fit results from the fitting procedure. The black histogram is the observed flux data. The black solid line is the estimated stellar continuum, which includes the intrinsic chromospheric emission lines of the star. The dashed black lines are the profiles of each absorption component, and the red line is the convolution of the assumed intrinsic stellar emission line folded through the interstellar absorption.}
\end{figure*}

\begin{figure*}[]
    \centerline{\includegraphics[trim={0 20cm 0 0},clip]{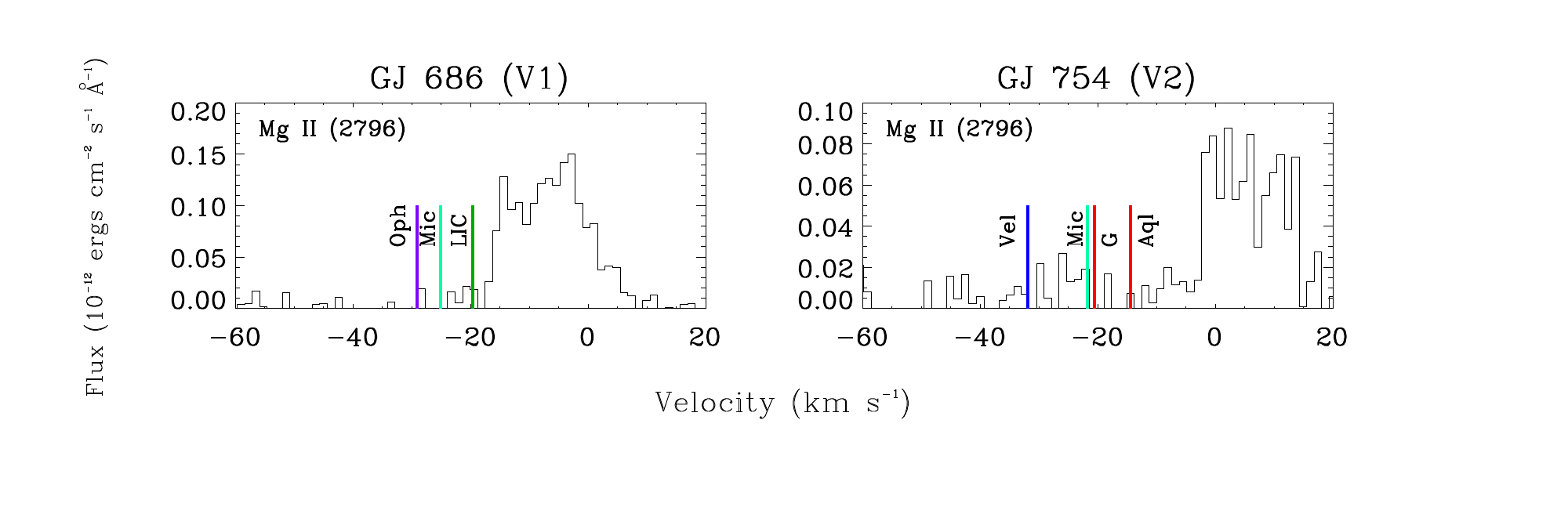}}
    \caption{Presented above is the \ion{Mg}{2} $\lambda$2796 line for both GJ 686 and GJ 754. The black histogram is the observed flux data plotted in velocity space. The colored vertical lines indicate velocities of the LISM clouds predicted to lie within $10^\circ$ of the star \citep{RedfieldLinsky2008}. The purple line is the Oph cloud, cyan is the Mic cloud, green is the LIC, blue is the Vel cloud, orange is the G cloud, and red is the Aql cloud.}
\end{figure*}

\begin{deluxetable}{ccccc}[]
\tablecaption{Fit Parameters for \textit{Voyager 1} LISM components.}
\tablehead{\colhead{Ion} & \colhead{Component} & \colhead{$v$} & \colhead{$b$} & \colhead{$\log_{10} N$}\\
\colhead{} & \colhead{\#} & (km $s^{-1}$) & (km s$^{-1}$) & (cm$^{-2}$)\\
}
\startdata
\multicolumn{5}{c}{GJ 678.1A}\\
\cline{1-5}
\ion{D}{1}\tablenotemark{c} & 1 & $-29.88\pm0.88$ & 7.65 & $13.52\pm0.09$\\
& 2 & $-20.3\pm1.9$ & 9.70 & $13.30\pm0.11$\\
\ion{Mg}{2} & 1 & $-29.47\pm0.88$\tablenotemark{a} & $1.51\pm0.54$\tablenotemark{b} & $12.72\pm0.41$\tablenotemark{b}\\
& 2 & $-23.70\pm0.66$\tablenotemark{b} & $2.09\pm0.47$\tablenotemark{b} & $14.29\pm0.34$\tablenotemark{b}\\
\cline{1-5}
\multicolumn{5}{c}{GJ 686}\\
\cline{1-5}
\ion{D}{1}\tablenotemark{c} & 1 & $-19.90\pm1.96$ & $9.62\pm2.35$ & $13.58\pm0.32$\\
\enddata
\tablenotetext{a}{Errors on \ion{Mg}{2} component 1 $v$ for GJ 678.1A are the standard deviation.}
\tablenotetext{b}{Errors on \ion{Mg}{2} component 1 $b$ and $\log_{10} N$, and component 2 $v$, $b$, and $\log_{10} N$ for GJ 678.1A are the weighted mean uncertainties.}
\tablenotetext{c}{Errors on \ion{D}{1} for GJ 686 and GJ 678.1A are MC uncertainties.}
\end{deluxetable}

\begin{deluxetable}{ccccc}[ht]
\tablecaption{Fit parameters for\textit{Voyager 2} LISM components.}
\tablehead{\colhead{Ion} & \colhead{Component} & \colhead{$v$} & \colhead{$b$} & \colhead{$\log_{10} N$}\\
\colhead{} & \colhead{\#} & (km $s^{-1}$) & (km s$^{-1}$) & (cm$^{-2}$)\\
}
\startdata
\multicolumn{5}{c}{GJ 780}\\
\cline{1-5}
\ion{D}{1}\tablenotemark{c} & 1 & $-16.8\pm1.6$ & $6.0\pm1.2$ & $12.78\pm0.25$\\
& 2 & $-10.4\pm2.9$ & $7.1\pm1.6$ & $12.85\pm0.32$\\
\ion{C}{2} & 1 & $-18.88\pm0.17$ & $2.86\pm0.38$ & $14.11\pm0.26$\tablenotemark{b} \\
& 2 & $-9.56\pm0.41$\tablenotemark{b} & $5.39\pm0.60$\tablenotemark{b} & $13.98\pm0.06$ \\
\ion{C}{2}$^*$\tablenotemark{c} & 1 & \nodata & \nodata & $12.37\pm0.19$\\
& 2 & \nodata & \nodata & $11.28\pm0.62$\\
\ion{O}{1}\tablenotemark{c} & 1 & $-16.3\pm1.3$ & $2.26\pm0.83$ & $14.88\pm0.59$\\
& 2 & $-8.20\pm0.64$ & $2.62\pm0.59$ & $14.43\pm0.34$\\
\ion{Mg}{2} & 1 & $-15.91\pm0.26$\tablenotemark{a} & $3.39\pm0.59$\tablenotemark{a} & $12.88\pm0.01$\tablenotemark{b} \\
& 2 & $-9.24\pm0.11$\tablenotemark{b}  & $3.52\pm0.10$\tablenotemark{b}  & $12.61\pm0.02$\tablenotemark{a}\\
\cline{1-5}
\multicolumn{5}{c}{GJ 754}\\
\cline{1-5}
\ion{D}{1}\tablenotemark{c} & 1 & $-27.4\pm2.6$ & $5.3\pm1.3$ & $13.19\pm0.29$\\
& 2 & $-18.5\pm1.2$ & $9.47\pm0.45$ & $13.17\pm0.10$\\
\enddata
\tablenotetext{a}{Errors on \ion{Mg}{2} component 1 $v$ and $b$ and component 2 $\log_{10} N$ for GJ 780are the standard deviation.}
\tablenotetext{b}{Errors on \ion{C}{2} and \ion{Mg}{2} component 1 $\log_{10} N$ and component 2 $v$ and $b$ for GJ 780 are the weighted mean uncertainties.}
\tablenotetext{c}{Errors on \ion{D}{1}, \ion{C}{2}$^*$, and \ion{O}{1} for GJ780 and \ion{D}{1} for GJ 754 are MC uncertainties.}
\end{deluxetable}

\subsection{The Ly$\alpha$ Profile}

The hydrogen Ly$\alpha$ line ($\lambda$1215.6700), the fundamental transition of the most abundant element, is central to studying the LISM and heliosphere. This wavelength is easily observed by STIS, so we were able to obtain high-resolution spectra of far-UV wavelength bands containing lines of both \ion{H}{1} and \ion{D}{1} Ly$\alpha$ ($\lambda$1215.3394). Interstellar hydrogen and deuterium atoms produce absorption features against the stellar hydrogen Ly$\alpha$ line profile \citep{Wood2005a}. These features provide important information about the LISM and are almost always strong enough to be observed for any sight line because hydrogen is so abundant. The Ly$\alpha$ line can also be used to detect the heliosphere and astrospheres, the structures analogous to the heliosphere around nearby stars \citep{Wood2005b}. 

The \ion{H}{1} and \ion{D}{1} Ly$\alpha$ lines are both close multiplets. The optically thin \ion{D}{1} Ly$\alpha$ provides initial conditions for reconstructing the \ion{H}{1} line profile. The \ion{D}{1} absorption fit should indicate what the central wavelength and Doppler width would be for \ion{H}{1} \citep{Wood2005a}. The \ion{H}{1} absorption has the same centroid velocity as deuterium, and the two Doppler parameters are related by $b(\mathrm{H I}) \approx \sqrt{2}b(\mathrm{D I})$ when thermal broadening dominates over nonthermal broadening, which is usually the case. This equation allows us to compute a Voigt opacity profile, $\tau_\lambda$, for \ion{H}{1} based on various assumed values for hydrogen column density \citep{Wood2005a}. This profile can effectively reconstruct the wings but not the core of the \ion{H}{1} Ly$\alpha$ line when we multiply the data by $\exp(\tau_\lambda)$ because the line core is highly saturated.

We follow the same method as outlined in \cite{Wood2005a} in that we reconstruct the stellar line profile by using a polynomial fit to the wings to interpolate between them, or use the \ion{Mg}{2} \textit{h} and \textit{k} lines to estimate the central shape. We are able to use \ion{Mg}{2} because Ly$\alpha$ and the \ion{Mg}{2} lines are optically thick chromospheric lines that have similar profiles in the solar spectrum.

In order to complete the best fit to each star, we specify certain parameters individually. All four stars show Ly$\alpha$ geocoronal emission in the middle of the LISM absorption, and this absorption is removed for fitting purposes. In GJ 780, we used the \ion{Mg}{2} fit parameters to fix the velocity separation and column density of the two ISM components, assuming identical $b$ values for both. We found it impossible to fit the \ion{H}{1} profile for GJ 780 solely with ISM absorption, with the data showing clear signs of both astrospheric absorption on the blue side of the line and heliospheric absorption on the red side. An additional single absorption component is used in the fit to approximate the combined heliospheric (HS) and astrospheric (AS) absorption (see Table 5). We use a single combined HS/AS component to estimate the amount of excess absorption on either side of the line that the ISM absorption cannot account for. We avoid the proliferation of unnecessary free parameters given that the blending of the components within the \ion{H}{1} Ly$\alpha$ line is severe. Interpreting the excess absorption associated with the heliosphere and astrosphere requires the assistance of hydrodynamic models. For GJ 754, we fix the velocity separation and column density ratios to be consistent with the \ion{D}{1} fit, again assuming identical $b$ values for each component. Like GJ 780, we use the \ion{Mg}{2} fit parameters for GJ 678.1A to fix the velocity separation and column densities of the two ISM components.  For GJ 686, we smooth the data by a factor of two to try to improve the low S/N. The data are fit with a single ISM component. 

We note that some degree of disagreement between the individual fits to \ion{D}{1} and the simultaneous fit to \ion{H}{1} and \ion{D}{1} is expected. We use a full reconstruction of the stellar Ly$\alpha$ profile including \ion{H}{1} and \ion{D}{1} to constrain $b$ and $v$ for each line. In the individual \ion{D}{1} fit, we use instead a polynomial interpolation to reconstruct the shape of the continuum. This may be the cause of systematic differences with regard to Doppler widths. The addition of \ion{H}{1} in a simultaneous fit and systematic differences in how the \ion{D}{1}-only continuum was created preclude larger $b$ values. We note that the relationship between $b$(\ion{H}{1}) $= \sqrt{2}b$(\ion{D}{1}) will not be strictly true, as we expect a small nonthermal contribution to the Doppler widths \citep{Wood2005a} . For fits with multiple LISM components (GJ 678.1A, GJ 780, and GJ 754), we fix the velocity separation based on the \ion{Mg}{2} fits when available. Since we detect no \ion{Mg}{2} absorption in GJ 754, the velocity separation is estimated from the \ion{D}{1} fit results. Separating components in \ion{D}{1} or \ion{H}{1} is far more difficult than in \ion{Mg}{2}, and in the case of GJ 754 it cannot be done to better than $\pm5$ km s$^{-1}$ considering the likely systematic errors.

The resulting fit parameters are presented in Table 5, and the fits to all four stars are presented in Figures 3--4. 

\begin{deluxetable}{ccccc}[hb]
\tablecaption{Ly$\alpha$ \ion{D}{1} and \ion{H}{1} Simultaneous Fit Parameters}
\tablecolumns{5}
\tablewidth{0pc}
\tablehead{
\colhead{Star} & \colhead{Component} & \colhead{$v$} & \colhead{$b$} & \colhead{$\log_{10} N$}\\
 & & (km s$^{-1}$) & (km s$^{-1}$) & (cm$^{-2}$)
}
\startdata
\multicolumn{5}{c}{\textit{Voyager 1}}\\
\cline{1-5}
GJ 678.1A & 1 & $-32.8\pm0.7$ & $14.5\pm1.2$ & $17.00\pm0.01$\\
& 2 & $-27.1$ & $14.5$ & $18.57$\\
GJ 686 & 1 & $-20.5\pm1.1$ & $13.1\pm1.4$ & $18.28\pm0.02$\\
\cline{1-5}
\multicolumn{5}{c}{\textit{Voyager 2}}\\
\cline{1-5}
GJ 780 & 1 & $-15.6\pm0.2$ & $12.0\pm0.2$ & $17.820\pm0.005$\\
& 2 & $-9.0$ & $12.0$ & $17.550$\\
& HS/AS & $-17.7\pm0.5$ & $22.1\pm0.6$ & $15.43\pm0.10$\\
GJ 754 & 1 & $-27.7\pm0.6$ & $11.3\pm0.4$ & $18.16\pm0.01$\\
& 2 & $-15.1$ & $11.3$ & $17.62$\\
\enddata
\end{deluxetable}

\begin{figure*}[h]
	\centerline{\includegraphics[scale=0.8,trim={1.9cm 12cm 1.9cm 1.9cm},clip]{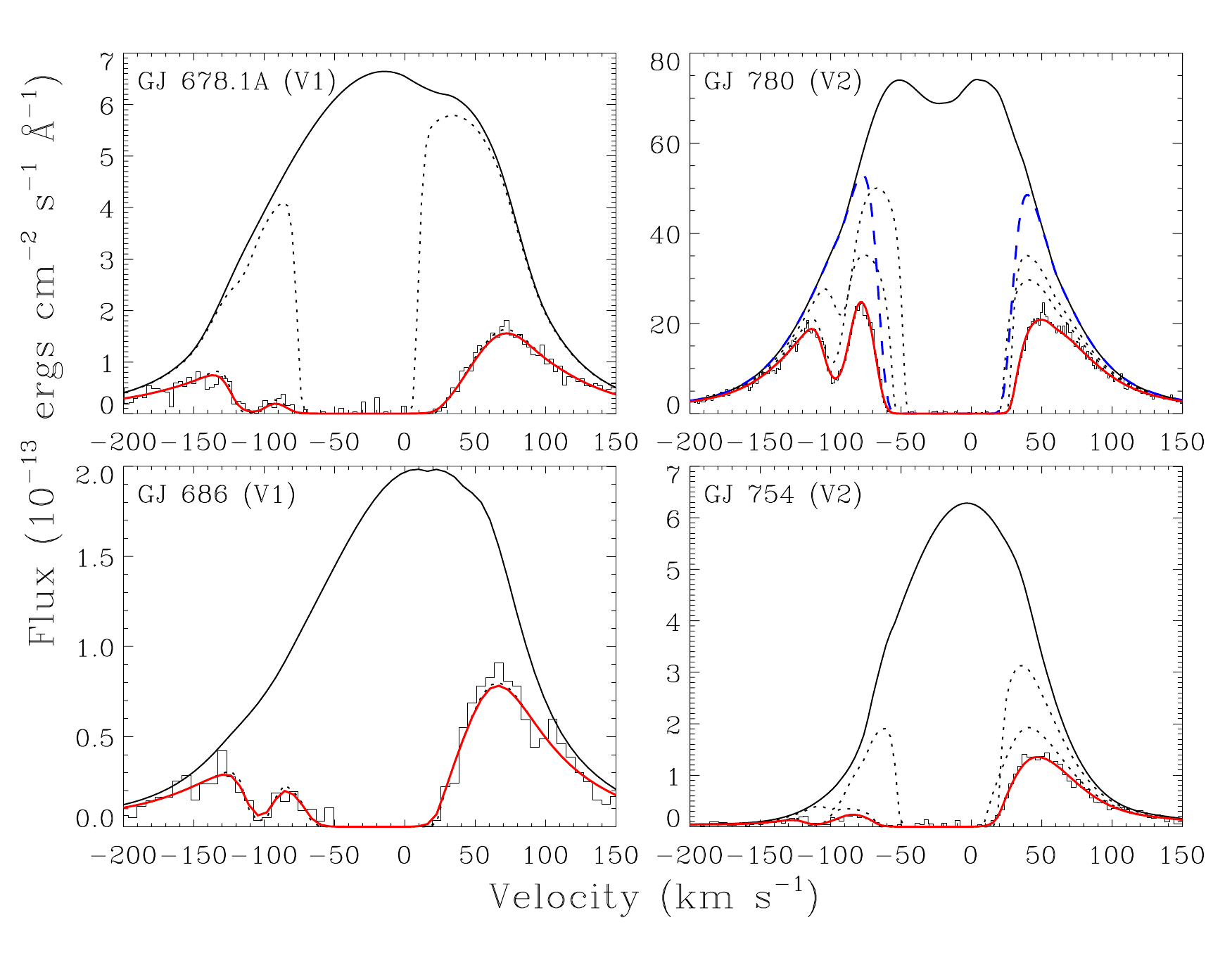}}
	\caption{The \ion{H}{1} Ly$\alpha$ spectra of our four target stars, plotted on a heliocentric
velocity scale, showing broad \ion{H}{1} absorption (centered at about $-20$ km s$^{-1}$) and narrow \ion{D}{1} absorption (centered at about $-100$ km s$^{-1}$) from the ISM superposed on the chromospheric emission line. The upper solid line is the reconstructed intrinsic stellar emission line profile. The dotted lines are the ISM components used to fit the data, and the blue dashed line (for GJ 780 only) is an extra absorption component used to approximate the combined heliospheric and astrospheric absorption in the fit. Thick red solid lines indicate the combined absorption of all components, after convolution with the instrumental line-spread function, which fits the data.}
\end{figure*}

\subsubsection{Integrated Flux Measurements}
We use the reconstructed stellar Ly$\alpha$ profile to measure integrated surface fluxes for our target stars. Characterizing the FUV ($1170-1700 \textrm{\AA}$) environment around exoplanet host stars is crucial for the study of exoplanet atmospheres and habitability. The integrated Ly$\alpha$ flux is a proxy for estimating the extreme-UV flux that heats exoplanetary upper atmospheres and drives mass loss \citep{Youngblood2016}. UV spectral behavior in M dwarfs is highly variable with order-of-magnitude increases during flares, which contributes to the strength of photolysis and photodissociation occurring in exoplanet atmospheres, and can account for the detection of biomarkers \citep{Miguel2015}. While none of our four target stars are currently known to host exoplanets, population studies indicate that they probably do (e.g. \cite{DressingCharbonneau2015}). Additionally, correlations between Ly$\alpha$ flux and other stellar emissions can aid the development of low-mass stellar atmospheric models \citep{Fontenla2016}. We compare our integrated Ly$\alpha$ surface flux measurements and Ly$\alpha$ and \ion{Mg}{2} surface flux ratios to those derived by \cite{Wood2005b}, \cite{France2013}, and \cite{Youngblood2016}. We present in Table 6 integrated Ly$\alpha$ and \ion{Mg}{2} flux measurements for all four target stars.

\begin{deluxetable*}{ccccc}[h]
\tablecaption{Ly$\alpha$ and \ion{Mg}{2} integrated fluxes}
\tablecolumns{5}
\tablewidth{0pc}
\tablehead{
\colhead{Star} & \colhead{$d$} & \colhead{$R$} & \colhead{$f$(Ly$\alpha$)} & \colhead{$f$(\ion{Mg}{2})}\\
& (pc) & ($R_\odot$) & (erg cm$^{-2}$ s$^{-1}$) & (erg cm$^{-2}$ s$^{-1}$)
}
\startdata
GJ 678.1A & $9.98$ & $0.543$\tablenotemark{a} & $5.6\times10^{-13}$ & $2.6\times10^{-13}$\\
GJ 686 & $8.09$ & $0.424$\tablenotemark{a} & $1.6\times10^{-13}$ & $2.1\times10^{-14}$\\
GJ 780 & $6.11$ & $1.223$\tablenotemark{b} & $5.2\times10^{-12}$ & $1.0\times10^{-11}$\\
GJ 754 & $5.71$ & $0.197$\tablenotemark{c} & $3.2\times10^{-13}$ & $1.5\times10^{-14}$\\
\enddata
\tablecomments{$^\mathrm{a}$\cite{Mann2015}, $^\mathrm{b}$\cite{Sousa2008},$^\mathrm{c}$\cite{Newton2017}}
\end{deluxetable*}

GJ 678.1A is an M1.0Ve \citep{Lepine2013} star with a ratio of the Ly$\alpha$ surface flux to the \ion{Mg}{2} \textit{h}$+$\textit{k} surface flux ($F$(Ly$\alpha$)$/F$(\ion{Mg}{2})) of 2.1, consistent with the best-fit line to M dwarf stars in Fig. 8 of \cite{Youngblood2016} and very similar to that of GJ 176, which has $F$(Ly$\alpha$)$/F$(\ion{Mg}{2}) = 1.9. We determine that the UV behavior of GJ 678.1A is consistent with other nearby M dwarfs of the same spectral type. 

GJ 686, which is an M1.5Ve \citep{Lepine2013} star, has a Ly$\alpha$ surface flux ($F_{\mathrm{Ly}\alpha} = 1.1\times10^5$ ergs cm$^{-2}$ s$^{-1}$) similar to that of later-type M dwarfs like GJ 876 (M3.5V; $F_{\mathrm{Ly}\alpha} = 1.2\times10^5$ erg cm$^{-2}$ s$^{-1}$, from \cite{Youngblood2016}). GJ 754 (M4V; \cite{Bonfils2013}) has a Ly$\alpha$ surface flux of $F_{\mathrm{Ly}\alpha} = 5.3\times10^5$ erg cm$^{-2}$ s$^{-1}$, somewhat larger than that measured by \cite{France2013} for GJ 876 (M4V; $F_{\mathrm{Ly}\alpha} = 3.6\times10^5$ erg cm$^{-2}$ s$^{-1}$). The integrated \ion{Mg}{2} fluxes for both stars are on order of magnitude smaller than that of GJ 678.1A, and produce small \ion{Mg}{2} surface fluxes. Therefore, both GJ 686 and GJ 754 have $F$(Ly$\alpha$)$/F$(\ion{Mg}{2})$>5$, and are inconsistent with the best-fit line for M dwarfs in Fig. 8 of \cite{Youngblood2016}. 

GJ 780 is a late G subgiant (G8IV) with an $F$(Ly$\alpha$)$/F$(\ion{Mg}{2}) ratio of 0.5, four times smaller than that for M dwarfs. In comparison to Fig.15b in \cite{Wood2005b}, we note that GJ 780 does correlate with the other listed G stars but is higher than the main-sequence best-fit line. 

\begin{figure*}[h]
	\centerline{\includegraphics[scale=0.9,trim={1.9cm 18cm 1.9cm 1.9cm},clip]{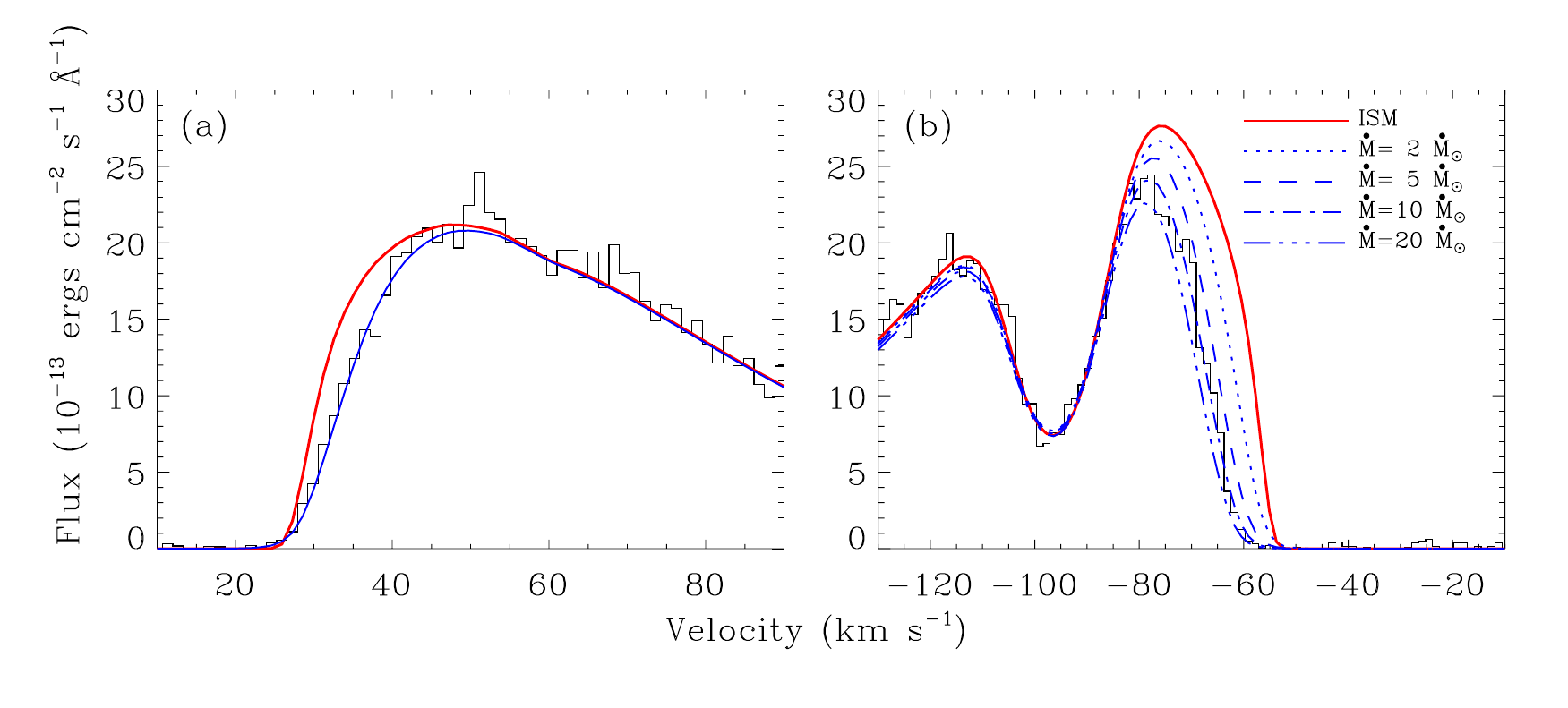}}
	\caption{(a) Close-up of the Ly$\alpha$ spectrum of GJ 780 from Figure 3, zooming in
on the red side of the \ion{H}{1} absorption. The red line is the total ISM absorption based on
the fit from Figure 3. The excess absorption is heliospheric absorption. The blue line shows
the absorption after including heliospheric absorption predicted by a hydrodynamic model
of the global heliosphere from \cite{Wood2000}. (b) Close-up of the GJ 780 Ly$\alpha$
spectrum, zooming in on the blue side of the \ion{H}{1} absorption. The red line again shows the
ISM absorption alone, with the excess absorption in this case being astrospheric absorption.
We show the predicted absorption of four models of the astrosphere, assuming different
stellar mass-loss rates of $\dot{M} = 2-20 \dot{M}_\odot$, with $\dot{M} = 10 \dot{M}_\odot$ providing the best fit to the
data.}
\end{figure*}
 
\subsubsection{Detection of Heliosphere and Astrosphere in GJ 780}
We detect excess absorption on both sides of the interstellar hydrogen absorption profile in GJ 780 (see Fig. 4), which indicates the presence of an astrosphere (absorption feature blueward of the hydrogen absorption; \cite{LinskyWood2014}) and a detection of the heliosphere (absorption redward of the interstellar hydrogen absorption; \cite{Wood2005b}).  For most sight lines, heliospheric Ly$\alpha$ absorption is dominated by the hydrogen wall (H-wall), the region outside the heliopause where interstellar \ion{H}{1} is heated, compressed, and decelerated \citep{Wood2007}.  Model predictions indicate that \ion{H}{1} deceleration within the H-wall relative to LISM flow is strongest in the upwind direction. The heliospheric absorption on the right side of the interstellar absorption is exactly the amount expected for this particular line of sight, about 58$^\circ$ from upwind \citep{RedfieldLinsky2008}. No other target showed evidence of heliospheric absorption, despite all being in the upwind direction. Even with higher S/N, we may not detect either type of absorption since all three stars have high interstellar \ion{H}{1} column densities that obscure the heliosphere and astrosphere absorption. Fig.13 in \cite{Wood2005a} shows that ISM column densities of $\log N$(H)$<18$ are generally necessary to narrow the ISM absorption enough to detect the heliospheric absorption signature.

Figure 4(a) shows the red side of the Ly$\alpha$ \ion{H}{1} absorption in GJ 780. The red line is the total ISM absorption based on Fig. 3, while the blue line shows the absorption after including heliospheric absorption. We use the hydrodynamic model of the heliosphere from \cite{Wood2000} that provides velocity distributions throughout the heliosphere, which we use to compute absorption profiles. We see that the \cite{Wood2000} estimate matches with the observed \ion{H}{1} profile. The absorption we detect comes from the outer heliosheath region that \textit{Voyager 2} will enter when it crosses the heliopause.

GJ 780 is only the 14th Ly$\alpha$ astrosphere detection based on \cite{Wood2005a}, \cite{Wood2005b}, and \cite{Wood2014}. With this detection, we estimate the mass-loss rate for GJ 780, since the amount of absorption is correlated with the strength of the stellar wind \citep{Wood2005b}. Measuring the mass-loss rate requires the ISM wind velocity seen by the star and the orientation of the astrosphere relative to the line of sight. We find that GJ 780 sees an ISM velocity of $v_{\mathrm{ISM}} = 29.4$ km s$^{-1}$, similar to that of $\epsilon$~Eri ($v_{\mathrm{ISM}} = 27$ km s$^{-1}$; \cite{Wood2005b}). Thus, we can use astrospheric models previously computed for $\epsilon$~Eri as models for the astrosphere of GJ 780.

In Figure 4(b), we show the predicted absorption of four models of the astrosphere, assuming different stellar mass-loss rates of $\dot{M} = 2 - 20\dot{M}_\odot$, where $\dot{M}_\odot \approx 2 \times 10^{-14}M_\odot$ yr$^{-1}$ \citep{Feldman1977}. We determine that the best-fit mass-loss rate for GJ 780 is $\dot{M} = 10\dot{M}_\odot$. For Sun-like stars, X-ray emission and stellar winds both originate in the stellar corona. \cite{Wood2005b} derived a power-law relation between mass-loss per unit surface area and stellar X-ray surface fluxes ($F_\mathrm{X}$). We calculate $F_\mathrm{X}$ for GJ 780 with the X-ray luminosity ($\log_{10} L_\mathrm{X} = 27.29$ erg s$^{-1}$)\footnote{https://www.hs.uni-hamburg.de/DE/For/Gal/Xgroup/nexxus/nstarpage.cgi?identifier=2201} and the stellar radius ($R = 1.223R_\odot$; \cite{Sousa2008}) and find $F_\mathrm{X}\approx 2.1\times 10^{4}$ erg cm$^{-2}$ s$^{-1}$. The high mass-loss rate estimated for GJ 780 would place it well above the $\dot{M}-F_\mathrm{X}$ relation defined by the main-sequence stars (\cite{Wood2005a}; \cite{Wood2014}), but GJ 780 is a subgiant rather than a main-sequence star.  The only other inactive subgiant with a measured wind, $\delta$~Eri (K0IV), also lies above the $\dot{M}-F_\mathrm{X}$ relation \citep{Wood2005a}.  Thus, the $\delta$~Eri and GJ 780 wind measurements suggest that inactive subgiants possess stronger winds than main-sequence stars, presumably due to lower surface gravity and lower surface escape speeds.

\section{Analysis of LISM Physical Parameters} \label{sec:analysis}
\subsection{Comparison with the LISM Dynamical Model}
\cite{RedfieldLinsky2008} developed a model of 15 distinct LISM clouds with bulk motion velocity vectors that explain the majority of LISM observations. An online LISM Dynamical Model Kinematic Calculator\footnote{http://lism.wesleyan.edu/LISMdynamics.html} identifies which LISM clouds are traversed for any given sight line and calculates the radial and transverse velocities for each of the 15 clouds. In Table 7, we present a comparison between our measured radial velocities and those from the clouds in \cite{RedfieldLinsky2008}. 

\begin{deluxetable*}{cccccc}[h]
\tablecaption{Comparison of Observed LISM Velocities with the \cite{RedfieldLinsky2008} LISM Dynamical Model}
 \tablecolumns{5}
\tablewidth{0pc} 
\tablehead{
\colhead{Star} & \colhead{Component} & \colhead{$v_{\mathrm{measured}}$} & \colhead{$v_{\mathrm{calculated}}$\tablenotemark{a}} & \colhead{Cloud} & \colhead{Alternate Cloud}\\
 & \colhead{No.} & (km s$^{-1}$) & (km s$^{-1}$) & \colhead{Name} & (km s$^{-1}$)
}
\startdata
\multicolumn{6}{c}{\textit{Voyager 1}}\\
\cline{1-6}
GJ 678.1A\tablenotemark{b} & 1 & $-31.5\pm2.4$\tablenotemark{c} & $-29.66\pm0.64$ & Oph & NGP ($-33.88\pm1.34$) \\
& 2 & $-23.9\pm2.4$\tablenotemark{c} & $-22.15\pm0.99$ & LIC &-\\
GJ 686 & 1 & $-20.5\pm1.1$ &  $-19.54\pm1.09$ & LIC &-\\
\cline{1-6}
\multicolumn{6}{c}{\textit{Voyager 2}}\\
\cline{1-6}
GJ 780\tablenotemark{b} & 1 & $-17.2\pm1.5$\tablenotemark{c} & $-18.71\pm1.09$ & Vel &- \\
& 2 & $-9.23\pm0.58$\tablenotemark{c} & $-8.62\pm0.60$ & Dor &- \\
GJ 754 & 1 & $-27.7\pm0.6$ & $-31.89\pm1.52$ & Vel & Mic ($-21.68\pm1.19$) \\
& 2 & $-15.1$ & $-14.36\pm1.04$ & Aql &- \\
\enddata

\tablenotetext{a}{\cite{RedfieldLinsky2008}}
\tablenotetext{b}{$v_{\mathrm{measured}}$ is the weighted mean of the velocities found for each ion, excluding the individual fit to \ion{D}{1}.}
\tablenotetext{c}{Errors are the standard deviation.}
\end{deluxetable*}

One important feature of the Kinematic Calculator is that it not only returns the projected velocity vectors for all 15 clouds,but also
lists which clouds are traversed by the sight line, which clouds are within $<$20$^\circ$ of the sight line, and which clouds are far
($>$20$^\circ$) from the sight line. We predict which clouds correspond to each ISM absorption component based on both the
velocity vectors and whether a sight line traverses a cloud. If a second cloud is a possible match to the component velocity, it is listed under ``Alternative Explanation." We calculate $v_{\mathrm{measured}}$ by finding the weighted mean of the velocities for each ion, but we exclude the velocities generated from the individual fits to \ion{D}{1}. The velocities resulting from the fits to Ly$\alpha$ are simultaneous fits to both \ion{D}{1} and \ion{H}{1}, and therefore yield more robust measurements of physical properties.

\subsubsection{The \textit{Voyager 1} sight line}
We find that the velocities for GJ 678.1A along the \textit{Voyager 1}
sight line match up well with two known ISM cloud velocities. GJ 678.1A has the smallest angular separation ($\Delta\theta = 8^\circ.1$) from the \textit{Voyager 1} sight line, indicating a strong likelihood that whatever clouds we identify along its path will intersect with the path of \textit{Voyager 1}. The first absorption component has a velocity of $-31.5\pm2.4$ km s$^{-1}$, which is close to the predicted velocity for the Oph cloud ($v_r = -29.66\pm0.64$ km s$^{-1}$). The NGP cloud ($v_r = -33.88\pm1.34$ km s$^{-1}$) passes near ($<20^\circ$) to the GJ 678.1A line of sight and also provides a close match to our calculated velocity. 

The second component, with an average velocity of $-23.9\pm2.4$ km s$^{-1}$, is $<$1$\sigma$ away from the velocity of the Local Interstellar Cloud (LIC), which has $v = -22.15\pm0.99$ km s$^{-1}$. We believe that the observed component velocity matches up well with the calculated velocity for the LIC because it is known that LIC material surrounds the solar system \citep{RedfieldLinsky2000, McComas2015} and hence would expect it to be present. 

We find that our measured velocity for GJ 686 ($-20.5\pm1.1$ km s$^{-1}$) matches well with the \cite{RedfieldLinsky2008} calculated velocity for the LIC ($-19.54\pm1.09$ km s$^{-1}$). Since the LIC has the largest surface area of any LISM cloud, we expect stars close to the \textit{Voyager 1} sight line to display LISM absorption from the LIC. 

\textit{Voyager 1 itinerary}: At its current speed ($\sim$3.6 au yr$^{-1}$), \textit{Voyager 1} will need to travel for $\sim$130 yr before it reaches 600 au, where the neutral hydrogen density is predicted by \cite{Zank2013} to be $0.20$ cm$^{-3}$, the level anticipated for the LIC. At this distance, we anticipate that \textit{Voyager 1} will have entered the ``pristine" ISM. To provide a rough estimate of the path length of each cloud that the \textit{Voyager} spacecraft will pass through, we assume a cloud density $n$(H) $\sim$ 0.2 cm$^{-3}$ \citep{Frisch2011}. We calculate the path length for each cloud by dividing our observed $N$(H) by the cloud density $n$(H). We find, as predicted by \cite{RedfieldLinsky2008}, that the path length through the LIC for \textit{Voyager 1} is the largest. Its greatest path length ($\sim 6$ pc) along this sight line is in the direction of GJ 678.1A, while toward GJ 686 its path length is $\sim 3$ pc. 

\cite{RedfieldLinsky2008} found that the closest star displaying LISM absorption from the Oph cloud, 70 Oph, is at a distance of about 5.1 pc. There are two possibilities: either the LIC and the Oph clouds overlap in front of 70 Oph at 5.1 pc, or the Oph cloud may be diffuse and concentrated further toward 10 pc. The closest star in \cite{RedfieldLinsky2008} displaying LISM absorption from the NGP cloud is 61 Vir at a distance of 8.5 pc. Since GJ 678.1A is 9.98 pc away, the NGP cloud, if present, could be located toward the end of the sight line to the target star. In the direction of GJ 686, we see only LISM absorption from the LIC. However, the \cite{RedfieldLinsky2008} kinematic model indicates that this sight line directly traverses the Oph and Mic clouds. With higher-S/N spectra for this target, we may have been able to observe absorption from these clouds. Regardless, the LISM environment through which \textit{Voyager 1} continues to travel will be increasingly dominated by the LIC. 

We conclude that \textit{Voyager 1}, after reaching pristine ISM, will travel through the LIC and may enter the Oph and NGP clouds for a short distance.

\subsubsection{The \textit{Voyager 2} sight line}
Despite falling within 15$^\circ$ of the direct \textit{Voyager 2}
line of sight, the sight lines to GJ 780 and GJ 754 are actually greater than 20$^\circ$ apart. This large separation between the two targets lends some uncertainty to which clouds occur along the path of \textit{Voyager 2}. 

According to \cite{RedfieldLinsky2008}, the sight line to GJ 780 directly traverses the Vel
cloud. With a velocity of $-17.2\pm1.3$ km s$^{-1}$, the first component
of GJ 780 matches closely (within $1\sigma$) with the calculated velocity of the Vel cloud ($v_r = -18.71\pm1.09$ km s$^{-1}$). No other LISM cloud can account for the measured velocity.

The second observed absorption component of GJ 780 also appears consistent with the Dor cloud, which passes near
($<$20$^\circ$) the line of sight with a calculated velocity of
$-8.62\pm0.60$ km s$^{-1}$. We find that the second component has
a measured velocity of $-9.23\pm0.58$ km s$^{-1}$, consistent with the velocity of the Dor cloud to
1$\sigma$. No other cloud provides a sensible match. 

The velocities of the two components of GJ 754 are more complicated, as the observed absorption velocities can
be attributed to several LISM clouds. The sight line to GJ 754 directly traverses the Aql cloud and passes within 20$^\circ$ of the G, Mic, and Vel clouds. We observe that the first absorption component has a velocity of $-27.7\pm0.6$ km s$^{-1}$, which is not consistent with the calculated velocity of the Aql cloud. It is, however, within 3$\sigma$ away from the velocity of the Vel cloud ($v_r = -31.89\pm1.52$ km s$^{-1}$). Additional absorption due to the Mic cloud ($v_r = -21.68\pm1.19$ km s$^{-1}$) may account for this discrepancy. If we had ISM absorption in the \ion{Mg}{2} \textit{h} and \textit{k} lines, we may have been able to better constrain this component's measured velocity. 

The second component of GJ 754 has a velocity of $-15.1$ km s$^{-1}$, which is $<$1$\sigma$ from that of the Aql cloud ($-14.36\pm1.04$ km s$^{-1}$). However, we cannot definitively rule out some absorption from the G cloud because it passes within $20^\circ$ of the line of sight to GJ 754 and is the next-largest cloud after the LIC.

\textit{Voyager 2 itinerary}: The \textit{Voyager 2} sight line, to both target stars, is slightly more complicated. For GJ 780, we definitively see LISM absorption from the Vel and Dor clouds, two of the most distant clouds in the \cite{RedfieldLinsky2008} sample. GJ 780 is only 6 pc from Earth, meaning that one or both clouds are observable at closer distances to the closest stars observed by \cite{RedfieldLinsky2008} with the cloud's absorption. Along this sight line, the Dor cloud has an estimated path length of 0.57 pc, and the Vel cloud has a path length of $\sim 1$ pc. Based on the lack of additional absorption features in GJ 780, we conclude that there is a gap between the two clouds. While we do not directly detect LISM absorption from the LIC or G cloud along this line of sight, we expect that both clouds may be present, but with low column densities.

In Section 5.1.2 we note that GJ 780 and GJ 754 are more than $20^\circ$ apart. Because of this, the differences in observed LISM clouds are significant. For GJ 780 we only see LISM absorption from two clouds, while for GJ 754 we see two LISM absorption features that arise from four possible clouds: G, Aql, Mic, and Vel. The G cloud is the next-largest and next-closest LISM cloud after the LIC \citep{RedfieldLinsky2008}. However, its calculated velocity ($-20.55\pm1.51$ km s$^{-1}$) is far from our observed velocities. Therefore, we find that it is unlikely that \textit{Voyager 2} will traverse the G cloud. The path lengths of the three additional clouds along this sight line are difficult to calculate, since there may be significant overlap between them. GJ 754 has the shortest sight line at a distance of 5.9 pc. Approximately 2.3 pc of that distance could be filled by the Mic and Vel clouds. We conclude that \textit{Voyager 2} will first reach the Aql cloud and then the Mic and Vel clouds.

\subsection{Temperature and Turbulence}

From measured line widths of several ions, we derive estimates for the temperature and turbulence of the ISM along both sight lines. The equation that illustrates the relationship between Doppler parameter ($b$ [km s$^{-1}$]), temperature ($T$ [K]), and turbulent velocity ($\xi$ [km s$^{-1}$]) is
\begin{equation}
    b^2 = \frac{2kT}m + \xi^2 = 0.01662\frac{T}A + \xi^2,
\end{equation}
where $k$ is the Boltzmann constant, $m$ is the mass of the observed ion, and $A$ is the atomic weight of the element \citep{RedfieldLinsky2004b}. Equation 1 can be solved with just two line width components, but the two ions must have very different atomic weights in order to yield accurate measurements. 

We present in Figure 5, for GJ 678.1A and GJ 780, visualizations of the temperature and turbulent velocity determination. We also present in Table 8 the calculated temperatures and turbulent velocities. For the \ion{H}{1} Ly$\alpha$ line Doppler parameters in GJ 678.1A and GJ 780 without uncertainties we assume a 20\% error.

\begin{figure*}[!]
   \centerline{\includegraphics[angle=90,origin=c,scale=0.43,trim={1.9cm 0 4.3cm 0},clip]{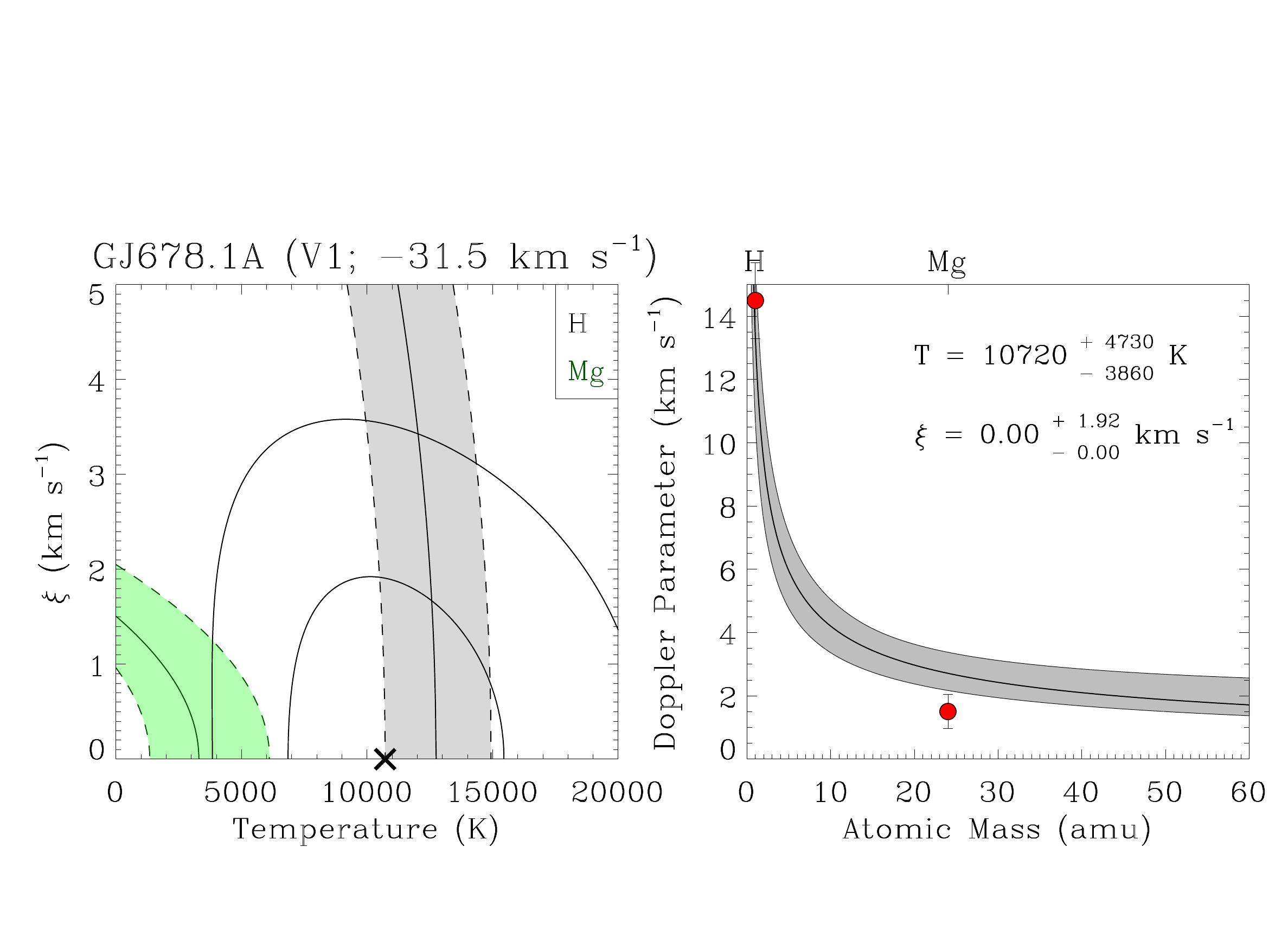}}
   \vspace{-2.5cm}
   \centerline{\includegraphics[angle=90,origin=c,scale=0.43,trim={1.9cm 0 4.3cm 0},clip]{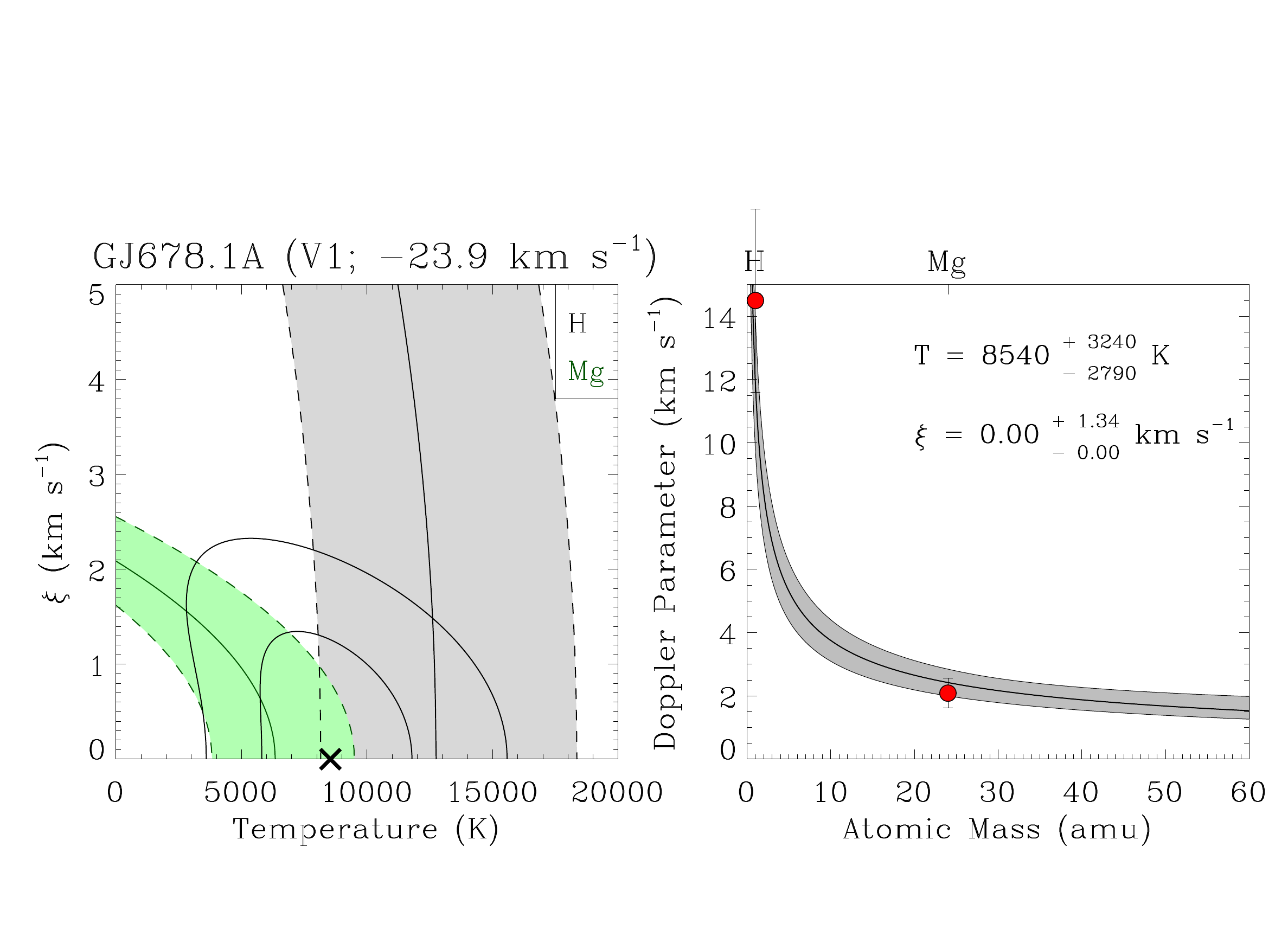}}
   \vspace{-2.5cm}
   \centerline{\includegraphics[angle=90,origin=c,scale=0.43,trim={1.9cm 0 4.3cm 0},clip]{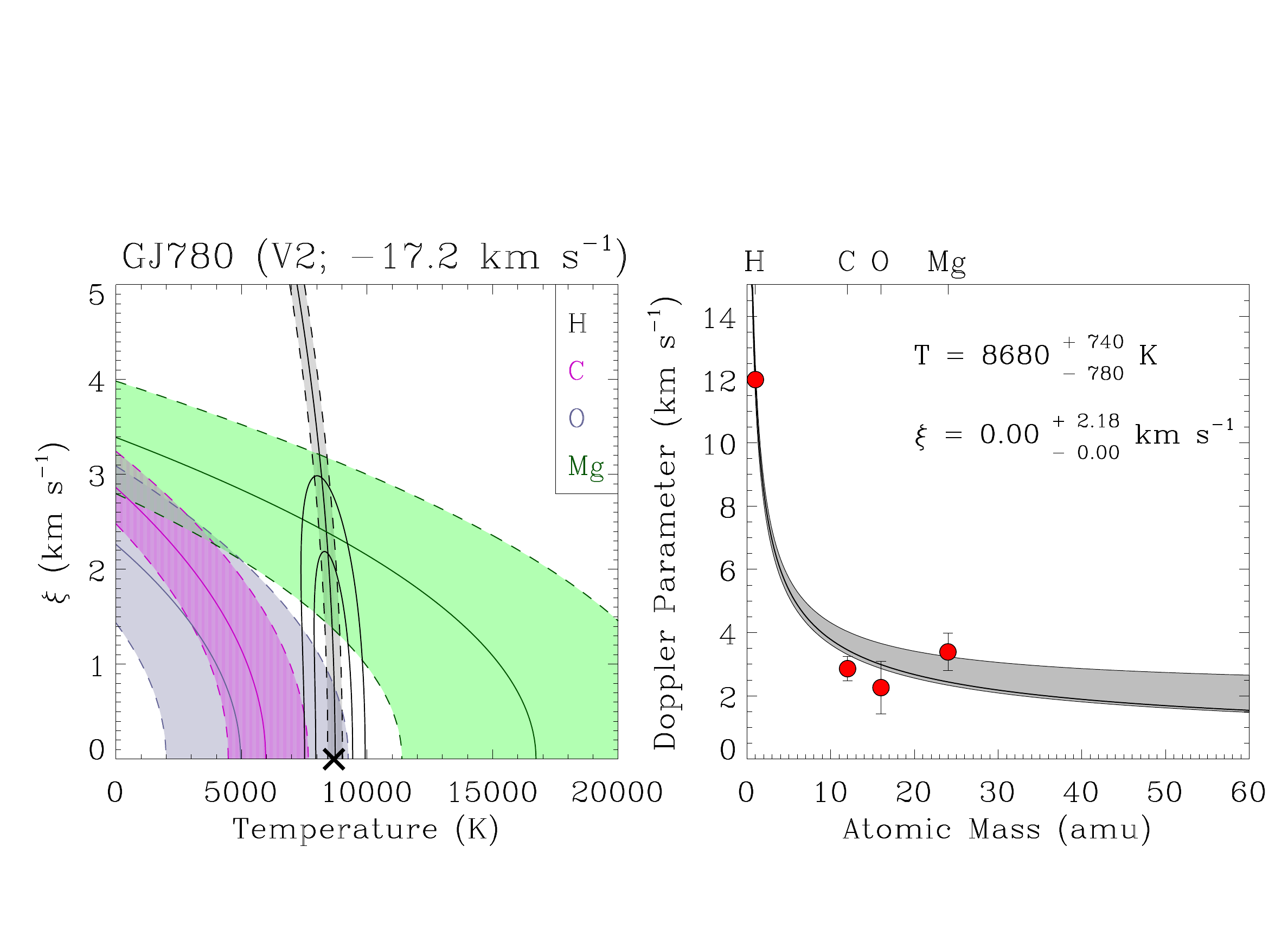}}
   \vspace{-2.5cm}
   \centerline{\includegraphics[angle=90,origin=c,scale=0.43,trim={1.9cm 0 4.3cm 0},clip]{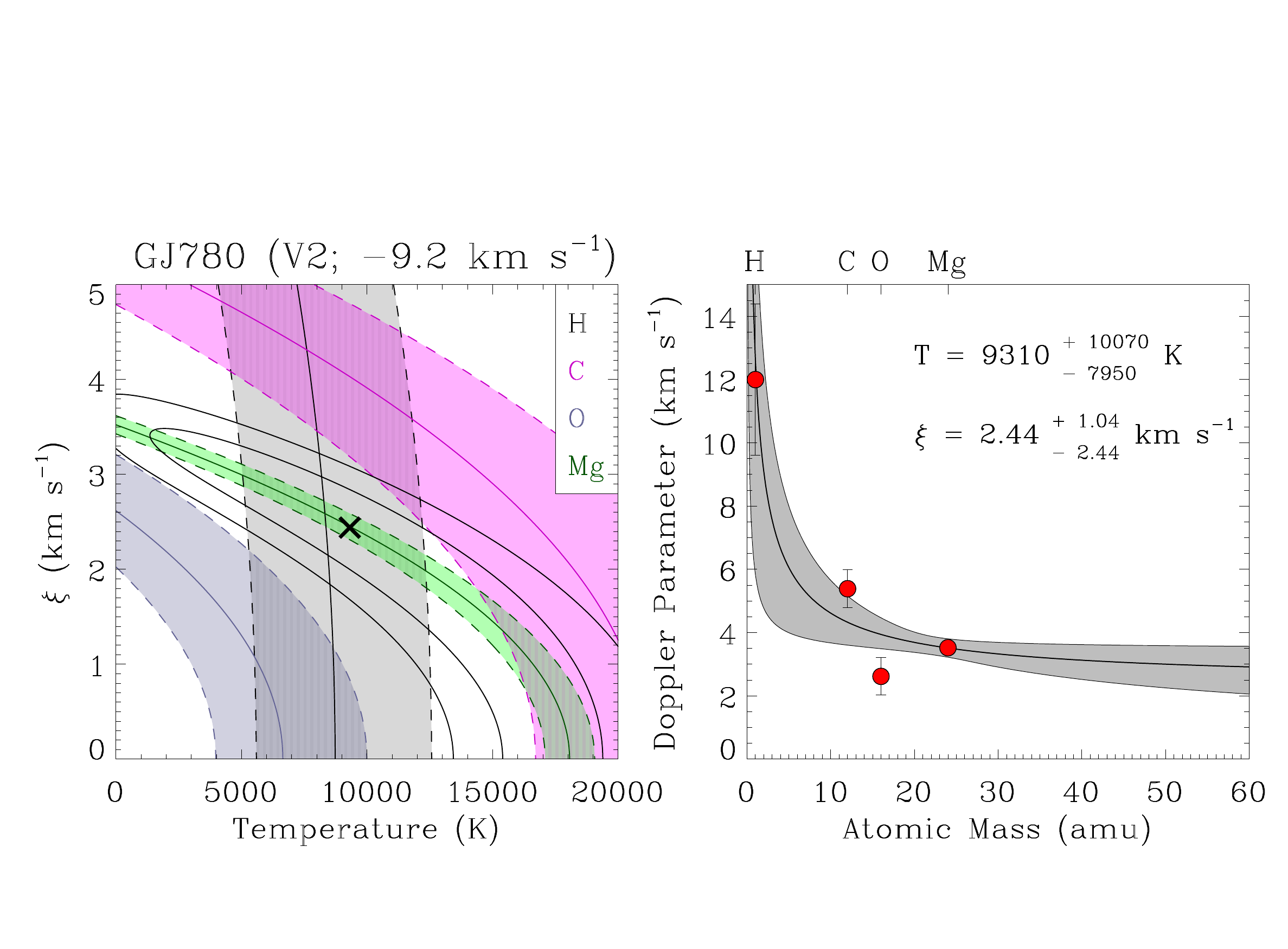}}
   \vspace{-3cm}
    \caption{Left panels: temperature versus turbulent velocity for the measured Doppler parameter of each ion. The colored dashed lines are $\pm$1$\sigma$ errors around the solid line, yielding an individual best-fit curve in the temperature-turbulence plane \citep{RedfieldLinsky2004b}. The black cross is the best-fit value of the temperature and turbulent velocity based on the measured Doppler parameters. The black contours are 1$\sigma$ and 2$\sigma$ errors for both values \citep{RedfieldLinsky2004b}. Right panels: Doppler parameters plotted against atomic mass. The red symbols are the observed Doppler parameter values and associated uncertainties from the data. The solid line is the best-fit curve given from Equation 1. The shaded regions include all fits within the 1$\sigma$ contours from the left panels.}
\end{figure*}

\begin{deluxetable}{ccccc}[ht]
\tablecolumns{5}
\tablewidth{0pc}
\tablecaption{Temperature and Turbulence Values for GJ 678.1A and GJ 780}
\tablehead{
\colhead{Gliese} & \colhead{Component} & \colhead{$v$} & \colhead{$T$} & \colhead{$\xi$}\\
\colhead{No.} & \colhead{No.} & (km s$^{-1}$) & (K) & (km s$^{-1}$)
}
\startdata
GJ 678.1A & 1 & $-31.5\pm2.4$ & $10720^{+4730}_{-3860}$ & $0.00^{+1.92}_{-0.00}$\\
& 2 & $-23.9\pm2.4$ & $8540^{+3240}_{-2790}$ & $0.00^{+1.34}_{-0.00}$\\
GJ 780 & 1 & $-17.2\pm1.5$ & $8680^{+740}_{-780}$ & $0.00^{+2.18}_{-0.00}$\\
 & 2 & $-9.23\pm0.58$ & $9310^{+10070}_{-7950}$ & $2.44^{+1.04}_{-2.44}$\\
\enddata
\end{deluxetable}

The temperature of the first GJ 678.1A LISM absorption component ($10720^{+4730}_{-3860}$ K) is high but may be reasonable given that the Oph cloud only has one other line of sight for which \cite{RedfieldLinsky2008} could obtain physical measurements. We provide the second determination of temperature and turbulence for a sight line to the Oph cloud. For the second LISM absorption component in GJ 678.1A, we report a temperature of $8540^{+3240}_{-2790}$ K. \cite{RedfieldLinsky2008} measure an average LIC temperature of 7500 $\pm$ 1300 K, and we conclude that our measurement is consistent with both the average and the temperatures measured along other LIC sight lines. Both absorption components have a median turbulent velocity of 0 km s$^{-1}$ with upper limits of $1.92$ km s$^{-1}$ and $1.34$ km s$^{-1}$, respectively. The average turbulences ($\xi_{\textrm{Oph}} = 3.3$ km s$^{-1}$ and $\xi_{\textrm{LIC}} = 1.62 \pm 0.75$ km s$^{-1}$) are within $3\sigma$ and $<1\sigma$ of our upper limit values.

We find a temperature of $8680^{+740}_{-780}$ K and a turbulent velocity of $0.00^{+2.18}_{-0.00}$ km s$^{-1}$ for the first LISM absorption feature in GJ 780. Our temperature is different from the 10,600 K temperature that \cite{RedfieldLinsky2008} found for the Vel cloud, but their temperature is uncertain and is based on only one sight line. Our upper limit on turbulent velocity is close to but smaller than that measured by \cite{RedfieldLinsky2008} ($\xi = 3.5$ km s$^{-1}$). The second LISM component, the Dor cloud, has a temperature of $9310^{+10070}_{-7950}$ K, which is much higher than the \citet{RedfieldLinsky2008} temperature of 7000K, also based on only one sight line. The large error bars are due to uncertainty in our measurement. We calculate a turbulent velocity ($\xi = 2.44^{+1.04}_{-2.44}$ km s$^{-1}$) that is slightly greater than 1$\sigma$ from the average Dor turbulence of 5.5 km s$^{-1}$. To better constrain temperature values for the identified clouds, we would need more observations of a wider range of ions.

The \textit{Voyager} spacecraft have and will experience a wide range of plasma temperatures as they continue to journey into the LISM. Between the heliopause and termination shock, temperatures can reach $2\times10^6$ K \citep{Richardson2009}. As \textit{Voyager 1} moves further into the LISM, the plasma temperature will continue to decrease. The temperature should also fluctuate as the spacecraft encounters different LISM component clouds, which have temperatures ranging between 5000 and 9900 K \citep{RedfieldLinsky2008}. Unfortunately, the Plasma Science (PLS) instrument on board \textit{Voyager 1} failed in 1980 and cannot directly measure temperatures. \textit{Voyager 2} is still in the heliosheath and, at its current speed, could reach the heliopause within the next 5 yr. Its PLS remains fully operational, and the spacecraft is able to measure the temperatures of the surrounding heliospheric plasma. Between its launch in 1980 and 2005, \textit{Voyager 2} measured interplanetary plasma temperatures between 3000 and 30,000 K \citep{Richardson2009}. After the spacecraft crossed the termination shock in 2007 at a distance of 84 au, there was a sharp jump in plasma temperature to over $10^5$ K, with temperatures varying between 20,000 and 200,000 K throughout the heliosheath \citep{Richardson2008, Richardson2009}. Once \textit{Voyager 2} reaches the LISM, it will observe an environment similar to the one \textit{Voyager 1} is in currently: first, a short distance through the LIC, with $T=7500\pm1300$ K and $\xi = 1.62\pm0.75$ km s$^{-1}$, and then the G cloud with $T = 5500\pm400$ K and $\xi = 2.2\pm1.1$ km s$^{-1}$ \citep{RedfieldLinsky2008}.

\subsection{Depletion}
In the 1990s, as it flew past Jupiter, \textit{Ulysses} confirmed the presence of interstellar dust grains within the solar system \citep{Frisch2009}. The dust, as expected, traveled with a speed and direction very similar to that of neutral interstellar hydrogen and helium gas. Since the interstellar clouds in the LISM were expected to be ``diffuse clouds" consisting largely of neutral hydrogen with low extinction coefficients, it was expected that the Local Cavity would be the same \citep{Draine2009}. Spectroscopic features observed in the infrared show silicate absorption, consistent with the assumption that most interstellar dust is composed of silicates or carbonates.

Depletion occurs when heavy elements have gas phase abundances with less than the expected cosmic abundances, presumably due to incorporation into dust. Depletion is calculated from
\begin{equation}
\log_{10}(X_{\mathrm{gas}}/\mathrm{H}) = \log_{10}(N(X)/N(\mathrm{H}))-\log_{10}(X/\mathrm{H})_{\odot},
\end{equation}
where $X$ denotes the heavier element, $N(X)$ is the column density of that element, and H is hydrogen \citep{Jenkins2009}. Depletion is determined by first calculating the ISM abundances and then subtracting the solar abundances \citep{Wood2002}. We adopt solar abundances ($\log_{10} X_\odot$) and the associated uncertainties from \citet{Asplund2009}. 

We derive $\log_{10}(X_{\mathrm{gas}}/\mathrm{H})$ for \ion{Mg}{2}, \ion{C}{2}, and \ion{O}{1} given the measured log column densities ($\log_{10} N(X)_\mathrm{gas}$ in Table 9) for each ion. In Table 9 we list the \ion{Mg}{2} depletions toward GJ 678.1A and the \ion{Mg}{2}, \ion{C}{2}, and \ion{O}{1} depletions toward GJ 780. The errors in the results from Equation 2 are calculated by error propagation based on the uncertainties on $\log_{10} N(X)$ and $\log_{10} X_\odot$.

\begin{deluxetable}{cccccc}[h!]
\tablecaption{Abundance and Depletion Values for GJ 678.1A and GJ 780}
\tablehead{
\colhead{Ion} & \colhead{Component} & \colhead{$\log_{10} N(X)_{\mathrm{gas}}$} & \colhead{$\log_{10} N$(H)} & \colhead{$\log_{10} X_\odot$\tablenotemark{a}} & \colhead{$\log_{10} X_\mathrm{gas}$/H}\\
\colhead{} & \colhead{No.} & ($\log_{10}$ cm$^{-2}$) & ($\log_{10}$ cm$^{-2}$) & \colhead{} & \colhead{}
}
\startdata
\multicolumn{6}{c}{GJ 678.1A}\\
\cline{1-6}
\ion{Mg}{2} & 1 & $12.72\pm0.41$ & $17.00\pm0.01$ & $7.60\pm0.04$ & $0.123\pm0.004$\\
 & 2 & $14.29\pm0.34$ & $18.57$ & \nodata & $0.115\pm0.004$\tablenotemark{b}\\
 \cline{1-6}
\multicolumn{6}{c}{GJ 780}\\
\cline{1-6}
\ion{C}{2} & 1 & $14.11\pm0.26$ & $17.82\pm0.005$ & $8.43\pm0.05$ & $-0.117\pm0.006$\\
& 2 & $13.98\pm0.06$ & $17.550$ & \nodata & $0.002\pm0.006$\tablenotemark{b}\\
\ion{O}{1} & 1 & $14.62\pm0.14$ & $17.82\pm0.005$ & $8.69\pm0.05$ & $0.114\pm0.007$\\
 & 2 & $14.03\pm0.57$ & $17.550$ & \nodata & $-0.209\pm0.007$\tablenotemark{b}\\
\ion{Mg}{2} & 1 & $12.88\pm0.01$ & $17.82\pm0.005$ & $7.60\pm0.04$ & $-0.537\pm0.004$\\
& 2 & $12.61\pm0.02$ & $17.550$ & \nodata & $-0.537\pm0.004$\tablenotemark{b}\\
\enddata
\tablenotetext{a}{With $\log_{10} H_\odot$ = 12.00.}
\tablenotetext{b}{Uncertainties were calculated by assuming 10\% error on $\log N$(H).}
\end{deluxetable}

We find that some of the observed depletions ($\log_{10}(X_{\mathrm{gas}}/\mathrm{H})$) indicate that the ions with negative gas column densities are locked in dust. The positive depletions represent that those ions remain in the gas phase. Our observations are in agreement with those of \cite{RedfieldLinsky2008}, who also measure both negative and positive \ion{Mg}{2} gas column densities for the LIC, Oph, Vel, and Dor clouds. In order to better constrain element abundances, we would need more ions like silicon or carbon in more sight lines. 

Dust grains exist over a wide range of sizes, with most mass concentrated below 0.25 $\mu$m \citep{Draine2009}. Since heavy elements can be concentrated in interstellar dust grains, and dust can move through interstellar gas, dust transport could account for variations in elemental abundances between LISM clouds \citep{Draine2004}. Furthermore, measurements of LISM elemental abundances and depletions can be used to estimate the total volume of silicate and carbonate dust \citep{Weingartner2001}. Based on the total estimated dust volume, \cite{Weingartner2001} determined the size distributions of dust. Silicate and carbonate dust grains have different size distributions, which can have a profound effect on any spacecraft traveling through the LISM, such as \textit{Voyager} and the proposed \textit{Breakthrough Starshot} mission. \cite{Hoang2017} found that ISM dust bombardment can erode a surface layer of $0.5$ mm and perhaps even modify the structure of a spacecraft. The Plasma Wave Subsystem (PWS) on board both \textit{Voyagers} detected small dust particles striking the spacecraft in interplanetary space \citep{Gurnett1997}. When micron-sized particles hit the spacecraft, they are vaporized, producing a a characteristic voltage pulse that is then detected in the wideband electric field waveform data \citep{Gurnett2015b}. \textit{Voyager 1} continues to measure three to seven impacts per hour \citep{Gurnett2015b}, with masses on the order of $10^{-10}$ to $10^{-11}$ g, placing the grains at the micron size. Due to the absence of latitudinal or radial gradients from observations taken in the early 2000s, \cite{Gurnett2005} concluded that \textit{Voyager 1} was already detecting grains of interstellar origin at 96.1 au. Therefore, as \textit{Voyager 1} moves further into the LISM, it will continue to see impacts by interstellar dust grains.

Some sight lines greater than 100 pc have much larger column densities and larger depletions than those we measure. \cite{Welsh2012} discuss long sight lines (100--200 pc) to white dwarfs, finding \ion{C}{2} and \ion{O}{1} log column densities of 14.6--14.8 cm$^{-2}$ and 14.5--15.2 cm$^{-2}$, respectively. These are considerably larger than the column densities measured for the lines of sight to GJ 678.1A and GJ 780. Additionally, the depletions for \ion{C}{2} are $-0.27$ to $-1.06$ and for \ion{O}{1} are $-0.51$ to $-1.07$ \citep{Welsh2012}. Again, these are much larger than for GJ 678.1A and GJ 780. The Local Cavity extends to roughly 100 pc from the Sun, and its shape is determined by the onset of significant \ion{Na}{1} absorption \citep{RedfieldLinsky2008}. Because the edge of the Local Cavity contains cold dense gas, there is more dust present, and consequently higher depletion. Therefore, local gas is warmer, has less dust, and has lower depletions than the gas near the edge of the Local Cavity.

\subsection{Electron density}
Because both \textit{Voyager} spacecraft are still capable of measuring electron densities with their PWS instruments, we compare our inferred electron densities from the \textit{HST} data to their {\em in situ} measurements. The detection of LISM absorption in lines of \ion{C}{2} and \ion{C}{2}$^*$ in the GJ 780 sight line enables a measurement of the electron density along the \textit{Voyager 2} line of sight. \cite{RedfieldFalcon2008} measured electron densities ($n_e$) using the ratio of the collisionally excited \ion{C}{2}$^*$ line column density to the column density of the \ion{C}{2} resonance line. The excited carbon line is a fine-structure doublet. Fine structure describes how spectral lines split into degenerate multiplets because of relativistic corrections leading to small shifts in energy (on the order of 10$^{-4} to 10^{-5}$eV). The presence of these fine-structure lines in LISM spectra gives information about the density of the absorbing medium \citep{BahcallWolf1968}.

The \ion{C}{2} resonance line at 1334.5323 $\textrm{\AA}$ corresponds to the transition from the ground state at J = $\frac{1}2$, while the collisionally excited \ion{C}{2}$^*$ doublet ($\lambda\lambda$1335.6627, 1335.7077) represents the transition from the excited state of the fine-structure doublet at J = $\frac{3}2$ \citep{BahcallWolf1968,RedfieldFalcon2008}. Since electron collisions are responsible for populating the excited state, we relate the ratio of the column densities to the electron density by
\begin{equation}
    \frac{N(\mathrm{C II}^*)}{N(\mathrm{C II})} = \frac{n_e C_{12}(T)}{A_{21}},
\end{equation}
where $N$(\ion{C}{2}$^*$) and $N$(\ion{C}{2}) are the column densities of the excited and resonance lines, respectively, and $A_{21} = 2.29\times 10^{-6}$s$^{-1}$ is the radiative de-excitation rate coefficient \citep{NussbaumerStorey1981}. The relation given in Equation 3 is derived from thermal equilibrium between the collisional excitation of the fine-structure doublet and the radiative de-excitation. The collision rate coefficient $C_{12}(T)$ can be written as
\begin{equation}
    C_{21}(T) = \frac{8.63\times 10^{-6}\Omega_{12}}{g_1T^{0.5}}\exp\bigg(-\frac{E_{12}}{kT}\bigg),
\end{equation}
where $g_1 = 2$ is the statistical weight of the ground state and $E_{12} = 1.31\times 10^{-14}$ erg is the energy of the transition. Following \cite{RedfieldFalcon2008}, we adopt the value of $\Omega_{12}=2.81$, and assume the weighted mean LISM temperature of $T = 6680$ K \citep{RedfieldLinsky2004b}. We present our calculations for electron density in Table 10.

\begin{deluxetable}{cccccc}[h!]
\tablewidth{0pc}
\tablecaption{GJ 780 election density values.}
\tablehead{
\colhead{Component} & \colhead{$v$} & \colhead{$\log_{10} N$(C II$^*$)} & \colhead{$\log_{10} N$(C II)} & \colhead{$T$} & \colhead{$n_e$}\\
\colhead{\#} & (km s$^{-1}$) & (cm$^{-2}$)& (cm$^{-2}$) & (K) & (cm$^{-3}$)
}
\startdata
1 & $-18.0\pm1.3$ & $12.37\pm0.19$ & $14.11\pm0.26$ & 6680 & $0.28\pm0.21$\\
2 & $-9.23\pm0.90$ & $11.28\pm0.62$ & $13.98\pm0.06$ & 6680 & $0.03\pm0.04$\\
\enddata
\end{deluxetable}

Our measurements of electron density for the \textit{Voyager 2} line of sight to GJ 780 clearly indicate the presence of the LISM. The first absorption component (Vel cloud) has an electron density of $0.28\pm0.21$ cm$^{-3}$. The large systematic error is due to the saturation of the \ion{C}{2} resonance line, which results from high \ion{C}{2} abundances and a high atomic oscillator strength \citep{Frisch2011}. This value of electron density is a factor of three greater than that measured by \textit{Voyager 1}, but it is consistent with previous predictions for electron density in the LISM. The higher density could suggest the presence of a high-density ISM cloud. 

The second absorption component (Dor cloud) has an electron density of 0.03 cm$^{-3}$, which is a factor of three smaller than that measured by \textit{Voyager 1}, but still a full order of magnitude greater than the heliosheath electron density. However, the upper limit of the uncertainty in the electron density is 0.07 cm$^{-3}$, which is both what was measured by \textit{Voyager 1} and close to the prediction by \cite{Frisch2011}. Once \textit{Voyager 2} crosses the heliopause within the next 5 yr and obtains new measurements of electron density, we will compare its future measurements with our derived values for electron density. 

In 2013 April, shortly after it crossed the heliopause, \textit{Voyager 1}'s PWS detected locally generated electron plasma oscillations at a frequency of $\sim$2.6 kHz \citep{Gurnett2013}. The \textit{Voyager} spacecraft last observed electron plasma oscillations in 2004 (\textit{V1}) and 2007 (\textit{V2}) when just upstream of the heliospheric termination shock. \cite{Gurnett2013} obtained a series of short samples of the electric field waveform and used Fourier analysis techniques to convert the waveforms into frequency vs. time spectrograms.

\cite{Gurnett2013} calculated that the observed frequency of 2.6 kHz corresponds to an electron density of $n_e = 0.08 $cm$^{-3}$. They also determined that the densities observed by \textit{Voyager 1} are increasing with radial distance at about 19\% per au. As of 2014, the electron density has increased slightly to about $0.09-0.11$ cm$^{-3}$ \citep{Gurnett2015}. Both of these observed electron densities are well within the range of remote-sensing measurements of plasma densities (0.06$-$0.21 cm$^{-3}$) in the LISM \citep{RedfieldFalcon2008}. Based on \textit{Voyager 2} PWS measurements out to 100 au, the electron densities in the heliosheath ($\sim$0.001$-$0.003 cm$^{-3}$) are at least an order of magnitude smaller than those in the LISM \citep{Gurnett2013}.

\section{Conclusions} \label{sec:conc}

We have acquired high-resolution \textit{HST}/STIS spectra along sight lines through the heliospheric and interstellar plasma that the \textit{Voyager} spacecraft are currently measuring, connecting two of NASA's highly successful and enduring missions. We demonstrate that the local ISM into which the \textit{Voyagers} are moving is a complex and rich environment.

\begin{enumerate}
\item We have created an ``Interstellar Road Map" for the \textit{Voyager} spacecraft. We use the \textit{HST} spectra to provide an overview of the ISM along the projected paths of the two \textit{Voyagers}. In the road map analogy, the spectra act as the basis for the map, providing a general idea of what lies along the path ahead. The \textit{Voyagers} themselves act as the ``street view," sending us valuable measurements of specific local physical properties. The one caveat is that the LISM is a dynamic structure $-$ it is constantly changing and evolving over time. Clouds that are along the \textit{Voyager} lines of sight now may not still be there when, in thousands of years, the spacecraft reach the interstellar space we have probed with \textit{HST}. Though our target stars are not exactly along the respective projected \textit{Voyager} lines of sight, they are all within 15$^\circ$ of it. 

\item We see multiple LISM cloud absorption toward the closest stars in those directions, and we confirm the presence of known interstellar clouds, including the LIC, along all four sight lines. In addition to the LIC, we see LISM absorption from the G, Oph, Vel, Aql, Mic, and Dor clouds. \textit{Voyager 1} will spend the next $\sim$130 yr in the outer heliosheath before it reaches pristine ISM material at 1000 au. The next $10^5$ years will be spent within the LIC. After \textit{Voyager 2} crosses the heliopause, presumably within the next 5 yr, we expect it to spend its next $10^5$ yr traveling a complex environment, including the G, Aql, Mic, and Vel clouds. We thereby provide an interstellar itinerary for the \textit{Voyager} spacecraft that predicts the environmental conditions for the next $10^5$ yr.

\item We report three new integrated intrinsic Ly$\alpha$ flux measurements for M dwarfs, adding to 11 calculated in prior studies by \cite{Linsky2013} and \cite{Youngblood2016}. We calculate the ratio of the intrinsic Ly$\alpha$ to \ion{Mg}{2} surface flux in GJ 678.1A and conclude that it is similar to that of other nearby M dwarf stars. We also calculate the same ratio for GJ 780 and determine it is higher than the derived bestfit for main-sequence G stars in Fig. 15b of \cite{Wood2005b}. This is as expected because GJ 780 is a subgiant and not a main-sequence star.

\item We detect the astrosphere around GJ 780 and estimate a mass-loss rate of $\dot{M} = 10\dot{M}_\odot$, suggesting that inactive subgiants have strong winds. The heliospheric detection in GJ 780 matches well with the model from \cite{Wood2000}. The absorption we detect from the heliosphere comes from the outer heliosheath region containing the hydrogen wall. This is the area through which \textit{Voyager 2} will enter after it crosses the heliopause.

\item We measure the physical properties, including temperature, turbulent velocity, and depletion, of the LISM clouds. We find that our temperatures are within 3$\sigma$ of the \cite{RedfieldLinsky2004b} weighted average for the LISM ($T = 6680\pm1490$ K). Our turbulent velocities are all less than $1\sigma$ from $\xi = 2.24\pm1.03$ km s$^{-1}$ derived by \cite{RedfieldLinsky2004b}. Based on \cite{RedfieldLinsky2008}, our depletion values are also typical. We measure electron density in the LISM and compare our values with those obtained by \textit{Voyager} \citep{Gurnett2013} and with the predicted range from \cite{Frisch2011}.

\end{enumerate}

Both \textit{HST} and the \textit{Voyagers} will be operational for at least the next few years, and so we must consider what additional goals we can accomplish with their remaining capabilities. We have taken observations from two of NASA's longest-enduring and incredibly successful missions and combined them to create an overview of local interstellar space along the paths of the two \textit{Voyager} spacecraft. While this is the first time these missions have been utilized together, hopefully it will not be the last.

\acknowledgements
We would like to acknowledge NASA HST Grant GO-13658 awarded by the Space Telescope Science Institute, which is operated by the Association of Universities for Research in Astronomy, Inc., for NASA, under contract NAS 5-26555.

\vspace{5mm}
\facilities{\textit{HST}(STIS)}

\end{document}